\documentclass[a4paper,fleqn]{cas-sc}

\usepackage[authoryear,longnamesfirst]{natbib}
\usepackage{amsfonts}
\usepackage{booktabs}
\usepackage{soul}
\usepackage[normalem]{ulem}
\usepackage[percent]{overpic}

\def\tsc#1{\csdef{#1}{\textsc{\lowercase{#1}}\xspace}}
\tsc{WGM}
\tsc{QE}

\newcommand{\avgg}[1]{\langle #1 \rangle_\gamma}
\newcommand{\avgn}[1]{\langle #1 \rangle}
\newcommand{\mode}[1]{\mbox{mode}( #1 )}
\newcommand{\trajds}{\mathcal{T}}
\newcommand{\trajdsD}{\trajds_{\downarrow}}
\newcommand{\trajdsU}{\trajds_{\uparrow}}
\newcommand{\xv}{\textbf{x}}
\newcommand{\vv}{\textbf{v}}
\newcommand{\prob}{\mathbb{P}}
\newcommand{\localization}{_hybrid}

\begin{document}
\let\WriteBookmarks\relax
\def\floatpagepagefraction{1}
\def\textpagefraction{.001}

\shorttitle{}

\title [mode = title]{High-statistics pedestrian dynamics on stairways
  and their probabilistic fundamental diagrams}

\author[1,2]{Caspar A.S. Pouw}[type=editor,auid=000,
bioid=1, orcid=0000-0002-3041-4533]
\cormark[1]
\ead{c.a.s.pouw@tue.nl}

\author[1,3]{Alessandro Corbetta}[auid=001]
\author[1,3]{Alessandro Gabbana}[auid=002]
\author[1]{Chiel van der Laan}[auid=003]
\author[1,3,4]{Federico Toschi}[auid=004]

\affiliation[1]{
  organization={
    Department of Applied Physics and Science Education, Eindhoven University of Technology
  },
  addressline={De Zaale},
  city={Eindhoven},
  postcode={5600 MB},
  state={Brabant},
  country={The Netherlands}
}

\affiliation[2]{
  organization={
    ProRail BV
  },
  addressline={Moreelsepark 2}, 
  city={Utrecht},
  postcode={3511EP}, 
  country={The Netherlands}
}

\affiliation[3]{
  organization={
    Eindhoven Artificial Intelligence Systems Institute, Eindhoven University of Technology
  },
  addressline={De Zaale},
  city={Eindhoven},
  postcode={5600 MB},
  state={Brabant},
  country={The Netherlands}
}

\affiliation[4]{
  organization={
    CNR-IAC
  },
  city={Rome},
  country={Italy}
}

\cortext[1]{Corresponding author}

\begin{abstract}
  Staircases play an essential role in crowd dynamics, allowing
  pedestrians to flow across large multi-level public facilities such
  as transportation hubs, shopping malls, and office buildings.
  Achieving a robust quantitative understanding of pedestrian behavior
  in these facilities is a key societal necessity. What makes this an
  outstanding scientific challenge is the extreme randomness intrinsic
  to pedestrian behavior. Any quantitative understanding necessarily
  needs to be probabilistic, including average dynamics and
  fluctuations. To this purpose, large-scale, real-life trajectory
  datasets are paramount.

  In this work, we analyze the data from an unprecedentedly
  high statistics year-long pedestrian tracking campaign, in which we
  anonymously collected millions of trajectories of pedestrians
  ascending and descending stairs within Eindhoven Central train
  station (The Netherlands). This has been possible thanks to a
  state-of-the-art, faster than real-time, computer vision approach hinged on 3D depth imaging, sensor fusion, and YOLOv7-based depth localization. 
  We consider both free-stream conditions, i.e.\ pedestrians walking
  in undisturbed, and trafficked conditions,
  unidirectional/bidirectional flows. We report on Eulerian fields
  (density, velocity and acceleration), showing how the walking
  dynamics changes when transitioning from stairs to landing. We then
  investigate the (mutual) positions of pedestrian as density changes,
  considering the crowd as a ``compressible'' physical medium. In
  particular, we show how pedestrians willingly opt to occupy fewer
  space than available, accepting a certain degree of
  compressibility. This is a non-trivial physical feature of
  pedestrian dynamics and we introduce a novel way to quantify this effect.
  As density increases, pedestrians strive to
  keep a minimum distance $d\approx 0.6\;$m (two treads of the
  staircase) from the person in front of them. Finally, we establish
  first-of-kind fully resolved probabilistic fundamental diagrams,
  where we model the pedestrian walking velocity as a mixture of a
  slow and fast-paced component (both in non-negligible percentages
  and with density-dependent characteristic fluctuations).  Notably,
  averages and modes of velocity distribution turn out to be
  substantially different.
  Our results, of which we include probabilistic parametrizations
  based on few variables, are key towards improved facility design and
  realistic simulation of pedestrians on staircases.
 
\end{abstract}

\begin{highlights}
\item High-statistics pedestrian dynamics on real-life staircases.
\item Highly-accurate anonymous pedestrian tracking via 3D depth
  imaging and state-of-the-art computer vision.
\item Phenomenological analysis via probabilistic fundamental diagrams and Eulerian fields.
\item Compressibility effects, position distributions
\end{highlights}

\begin{keywords}
  Crowd dynamics on stairways \sep Probabilistic fundamental diagrams \sep Eulerian fields \sep Compressibility and space occupancy \sep High-statistics
  trajectory dataset \sep Real-time pedestrian tracking
\end{keywords}

\maketitle

\section{Introduction}\label{sec:intro}
Staircases play an essential role in crowd dynamics, allowing pedestrians flows across large multi-level public facilities such as transfer stations, airport terminals, shopping malls, and office buildings. These facilities also often represent a primary source of crowd congestion and pedestrian accidents~\cite{jackson-1995-josr,templer-1995-staircase,prorail-report-2021,feliciani-ss-2023}. Established structural regulations, e.g. \cite{en2002eurocode}, impose performance requirements on facilities, e.g. in terms of capacity and safety, that need to be satisfied with a sufficiently high probability through the lifetime of the facility \citep{augusti-pem-2008}. To achieve such performance-based design, a deep phenomenological understanding of the crowd behavior within and around staircases, including averages, probabilities of fluctuations, and rare events is key. A profound understanding of the probabilistic landscape can therefore empower facility managers to gauge the frequency of dangerous events, e.g. bottlenecks, and thus anticipate potential capacity reductions.

In the past decades, the study of the dynamics of pedestrian crowds
has emerged as a multidisciplinary field across civil
engineering~\cite{vanamu-etrr-2017},
physics~\cite{corbetta-annurev-2023}, mathematical
modeling~\cite{cristiani2014multiscale}, computer
science~\cite{roland-inp-2016}, psychology~\cite{drury-cop-2020}, and
more~\cite{haghani-physa-2021}. Alongside its immediate societal
relevance, the study of pedestrian crowds shares deep connections with
the fundamental physics of active matter systems
\cite{corbetta-annurev-2023}.
Traditionally, crowd dynamics have been studied via small-scale laboratory setups, simulation models and/or surveys. Only during the last 10 years we established the technological capacity of performing measurement campaigns in real-life conditions. Pedestrian tracking in real-life opens up the possibility to record and study the dynamics of pedestrians on a significantly larger statistical scale, surpassing conventional (experimental) datasets by several orders of magnitude. High-statistics data is vital to characterize probabilities of fluctuations and quantify rare, potentially dangerous, events. However, in real-life contexts experimental parameters, such as crowd density, are not subject to deliberate specification, so measurements are inherently confined to the crowd dynamics manifested during the measurement campaign. Overhead depth sensing has emerged as an extremely robust option to this purpose, allowing for accuracy and respect of individual privacy~\cite{seer-trc-2014,corbetta-trp-2014,willems-sr-2020}.

In the context of pedestrian traffic, the main phenomenological
modelling elements are average macroscopic relationships among crowd
density and walking speed or flow. These average relations are
generally dubbed fundamental diagrams (FDs,
e.g.,~\cite{vanamu-etrr-2017}). For flat grounds, fundamental diagrams
have been extensively obtained in laboratory conditions considering,
among others, uni- and multi-directional flows and entailing different
populations (in terms, e.g., of age, geographical region,
competitiveness, \cite{cao-2018-physa,ren-2019-josm,subaih-cci-2019,ye-physa-2021}). FDs
are key for infrastructural design, supporting, e.g., the evaluation
of the capacity of a system. Yet, the dynamics of pedestrians is highly random due to, e.g.,
inter-subject and intra-subject variability, routing variability,
presence of groups. Fundamental diagrams simplify the highly random
dynamics of pedestrians with a single average. This reduction is
unavoidable when experimental data is limited to small-scale datasets,
for which statistical moments are far from convergence. On the other
hand, recent depth-based large-scale real-life campaigns have allowed
to integrate FDs with a resolved description of
fluctuations~\cite{gabbana-pnasn-2022, brscic-trp-2014}, and even
characterization of rare events happening, e.g., once in a thousand
pedestrians~\cite{corbetta-pre-2017,corbetta-pre-2018}.
Research on spatial distributions and relative distances in crowds, often referred to as proxemics, emerged with studies investigating the relative distances between people in small, mostly stationary, social activities \cite{hall-book-1966}. This research expanded to include spatial analysis of the walking behavior of pedestrian social groups, as demonstrated by~\cite{moussaid-plos-2012}, and later studied in more detail through the analysis of nearest neighbors in different flow types \cite{cao-jsm-2021}. These studies rely on experimental data, typically involving only a limited number of trajectories. While field studies do exist, e.g.~\cite{zanlungo-pre-2014} who studied the spatial distribution in groups of 2 or 3 pedestrians, the extraction of pedestrian trajectories from the recordings is manual and labor-intensive.
Large-statistics datasets open the possibility of systematic
analyses of the crowd as a fluid~\cite{hughes2003flow}, considering,
e.g., how the space is filled or how geometry influences the dynamics.
This knowledge is key towards a quantitative physical understanding of
the crowd behavior and realistic predictive simulations.

Pedestrian dynamics on staircases is a profoundly studied topic,
pioneered five decades ago by, among others \cite{fruin-book-1971,
  predtechenskii-report-1978}. They reported initial findings on the
free flow velocity and established correlations between walking speed
and crowd density.  Since then extensive research has been devoted
into the definition of FDs and in the study of the different factors
effecting them, such as type of infrastructures (considering slope,
tread-depth, riser height, ~\cite{burghardt-trpc-2013,wang-ss-2021},
or especially long staircases~\cite{kretz-ss-2008, ma-ss-2012,
  ronchi-report-2016, chen-ft-2018}), flow compositions
(unidirectional, bidirectional, ascending,
descending~\cite{chen-josm-2017,ye-ss-2023}), and cultural and
personal features (age, gender,
etc.~\cite{fujiyama-transed-2004}). Data has been collected via field
studies with limited volunteers, surveys, and small-scale
experiments~\cite{hankin-jors-1958, fruin-book-1971, lam-jte-2000,
  fujiyama-transed-2004, ye-trr-2008, peacock-ss-2012, qu-ss-2014,
  shi-physa-2021}. Leveraging on relatively limited datasets, of at
most few hundred data points, all previous studies have unavoidably
focused on deterministic average
behaviors~\cite{ye-ss-2023}. Conversely, for movement on stairs a
fully resolved characterization of fluctuations, complementing the
studies for level ground by~\cite{gabbana-pnasn-2022,
  brscic-trp-2014}, remains completely outstanding. The field of
proxemics is also represented in the literature that studied
pedestrian dynamics on staircases. \cite{burghardt-trpc-2013}, for
example, reports a topographical examination based on Eulerian fields
using an experimental dataset featuring young German student
descending a staircase. Building on this work,~\cite{ye-ss-2023}
revisited the same dataset, delving deeper into spatial analyses by
exploring lane formations and spatial
distributions. Additionally,~\cite{xie-jsm-2023} conducted a field
study focusing on parent-child pairs during a brief 505 second time
interval. With this work we aim to expand the literature with an
in-depth spatial analysis of a truly unbiased and high-statistics
trajectory dataset providing fundamental insights into pedestrian
behaviors and interactions.

\begin{table}[htb]
  \caption{Pedestrian dynamics on staircases in the literature. We report the horizontal walking speed (in free flow conditions) identified in previous studies, almost always measured in laboratory conditions, together with the magnitude of the dataset acquired. In this work, thanks to a real-life dataset including over 3 millions real-life trajectories we study at the statistical level pedestrian dynamics on staircases. Velocity distributions and fundamental diagrams turn out to have long tails due to the presence of multiple populations moving at different speed. This yield a substantial difference between the modes and the averages of the velocity distributions at all density levels. Besides, thanks to the large dataset we can investigate spatial fields of positions, velocities and accelerations and compressibility effects.
          }\label{tab:free-stream-vel}
  \resizebox{\columnwidth}{!}{
  \begin{tabular}{lllcccc} 
      \toprule
      {Source}  &  \multicolumn{2}{c}{Free flow mean speed (m/s)} & {Riser height } & {Tread depth } & {Slope} & \# Trajectories \\ 
                &  {Ascent} & {Descent}                 & (mm)            &  (mm)          & (°)         &            \\  
    \midrule
    \cite{fruin-book-1971}       & $0.57$          & $0.77         $ & 152 & 305 & 27.0 & $\mathcal{O}(10^3)$ \\
    \cite{fruin-book-1971}       & $0.51$          & $0.67         $ & 178 & 286 & 32.0 & $\mathcal{O}(10^3)$ \\
    \cite{frantzich-report-1996} & $0.51 \pm 0.10$ & $0.71 \pm 0.27$ & 205 & 225 & 42.3 & $\mathcal{O}(10^3)$ \\
    \cite{lam-jte-2000}          & $0.42         $ & $0.57         $ & 163 & 271 & 31.0 & $\mathcal{O}(10^3)$ \\
    \cite{fujiyama-transed-2004} & $0.68-0.77    $ & $0.80-0.91    $ & 152 & 332 & 24.6 & $\mathcal{O}(10  )$ \\
    \cite{kretz-ss-2008}         & $0.65         $ & $0.71         $ & 150 & 367 & 22.2 & $\mathcal{O}(10^2)$ \\
    \cite{peacock-ss-2012}       & -               & $0.45         $ & 186 & 238 & 38.0 & $\mathcal{O}(10^3)$ \\
    \cite{ma-ss-2012}            & -               & $0.55         $ & 150 & 280 & 28.2 & $\mathcal{O}(10^2)$ \\
    \cite{qu-ss-2014}            & -               & $0.63         $ & 300 & 140 & 25.0 & $\mathcal{O}(10^2)$ \\
    \cite{qu-ss-2014}            & $0.55         $ & -               & 330 & 157 & 26.1 & $\mathcal{O}(10^2)$ \\
    \cite{ronchi-report-2016}    & $0.62-0.75    $ & -               & 180 & 267 & 34.7 & $\mathcal{O}(10^2)$ \\
    \cite{chen-ft-2018}          & $0.50 \pm 0.17$ & $0.61 \pm 0.14$ & 160 & 260 & 32.0 & $\mathcal{O}(10^2)$ \\
    \cite{koster-pre-2019}       & $0.59         $ & $0.64-0.69    $ & 165 & 295 & 29.2 & $\mathcal{O}(10  )$ \\
    \hline\hline \\
    \multicolumn{3}{c}{\textbf{This work}} & \textbf{167} & \textbf{290} & \textbf{30.0} & \textbf{$\mathcal{O}(10^6)$} \\ \hline
    \textbf{Free flow speed - average} & $0.63$ & $0.73$ & &  &  & \\
    \textbf{Free flow speed - mode}    & $0.48 \pm 0.08 $ & $0.58 \pm 0.1  $ & & & & \\
    \textbf{Probabilistic fundamental diagrams}   & Figure~\ref{fig:fd-ps}a & Figure~\ref{fig:fd-ps}b & &  &  &  \\
       & Eq.~\eqref{eq:gauss-mixture-equal} and Eq.~\eqref{eq:fit-ascent} & Eq.~\eqref{eq:gauss-mixture-equal} and Eq.~\eqref{eq:fit-descent} & &  &  &  \\
    \textbf{Eulerian fields (positions/velocity/accelerations)} & \multicolumn{2}{c}{Figure~\ref{fig:heatmaps}} & & & \\
    \textbf{Space occupation and compressibility effects} & \multicolumn{2}{c}{Figures~\ref{fig:density-def},\ref{fig:nearest-neighbour-heatmap},\ref{fig:neighbor-distance}} & & & \\
    \bottomrule
  \end{tabular}
  }
\end{table}

In this work, we perform a thorough phenomenological analysis of the
statistical behavior of pedestrians moving on a staircase within
Eindhoven Central Train Station (The Netherlands). To this purpose, we
collect an unprecedented dataset consisting of over 3 millions
trajectories, during a year long real-life pedestrian tracking
campaign. We employ an anonymous depth-based overhead tracking system
(cf. previous works by the
authors~\cite{corbetta-pre-2017,corbetta-pre-2018,gabbana-pnasn-2022})
that, for the first time, we generalize to operate in all three
dimensions.

We study both microscopic and macroscopic features of crowd flows
observed from data: we analyze density patters and interpersonal
distance of pedestrians in diverse flow configurations, ranging from
dilute up to highly dense conditions. We analyze how pedestrians fill
the available space as their number increases: considering the physics
concept of compressibility factor, we show how pedestrians fill the
space, behaving differently from an ideal gas that would expand
occupying all the available space. In particular, at variance with an
ideal gas, we model how pedestrians operate filling all the available
space. Moreover, we investigate the relationship between density and
velocity under diverse flow conditions, including comparisons with
experimental and theoretical literature data
(cf. Table~\ref{tab:free-stream-vel}). Thanks to our high statistics
dataset we are able to present a complete parametrization of the
probability distribution function of the velocity, as a function of
density.

This work is structured as follows: in Sec.~\ref{sec:data-collection}
we describe our data acquisition campaign and measurement process.  We
detail the environmental setup and our depth sensors installation in
Sec.~\ref{sec:experimental-setup}. In Sec.~\ref{sec:computer-vision},
we report the processing pipeline which yields anonymous high-quality
pedestrian trajectories based on a stream of depth images. This
combines perspective corrections, sensor-fusion, state-of-the-art
machine-learning-based pedestrian localization and Lagrangian
tracking. We leverage specifically on a machine-learning based
tracking algorithm (YOLOv7,~\cite{yolov7-arxiv-2022}) which enables us
to achieve high quality localization even in the most crowded
scenarios. In Sec.~\ref{sec:density-vel}, we discuss the definition
and computation of macroscopic variables, such as velocity and
density, from trajectory data. In Sec.~\ref{sec:results} we present
the results of our analysis which are further organized as follows:
\begin{itemize}
\item In Sec.~\ref{sec:data-overview}, we provide an overview of the
  data, detailing the most common flow conditions on the staircase. We
  also evaluate spatial (Eulerian) distributions of positions,
  velocities and accelerations to quantify the usage. We then
  provide a quantitative modeling of the free-stream velocity
  probability distribution function.
\item In Sec.~\ref{sec:compressibility}, we analyze interactions
  between pedestrians, discussing crowd compressibility effects and
  the relative positioning of nearest neighbors at increasingly large
  values of the local density.
\item In Sec.~\ref{sec:probab-fund-diag} we establish a probabilistic
  fundamental diagram, where we describe the parametrization of the
  probability distribution function of the velocity, as a function of
  the local density.
\end{itemize}
Finally, in Sec.~\ref{sec:conclusions} we discuss and summarize main 
findings and future developments.

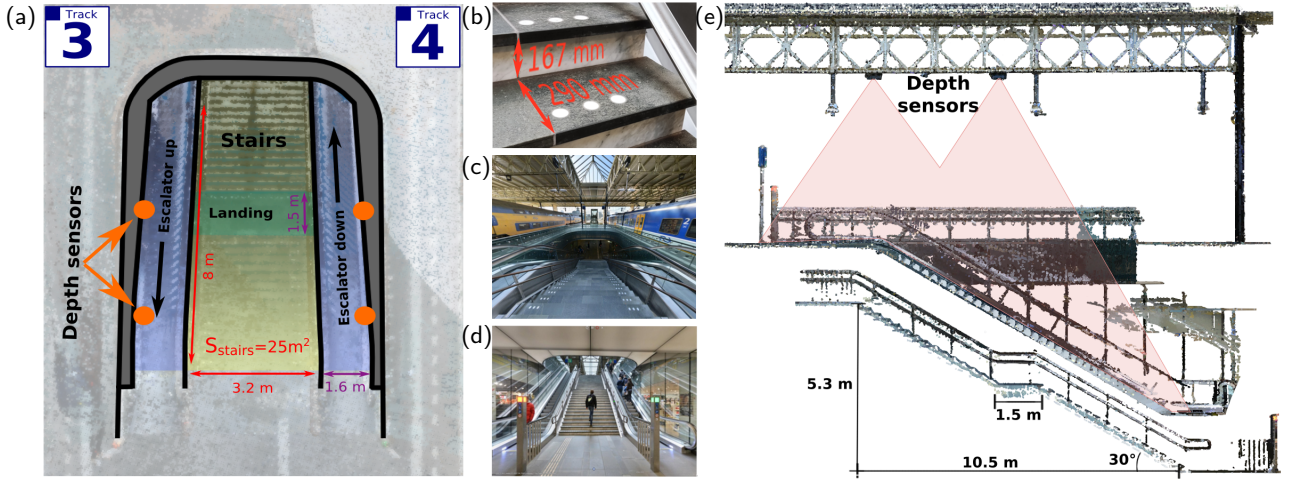
\begin{figure}[t]
  \centering

  \begin{overpic}[width=\linewidth]{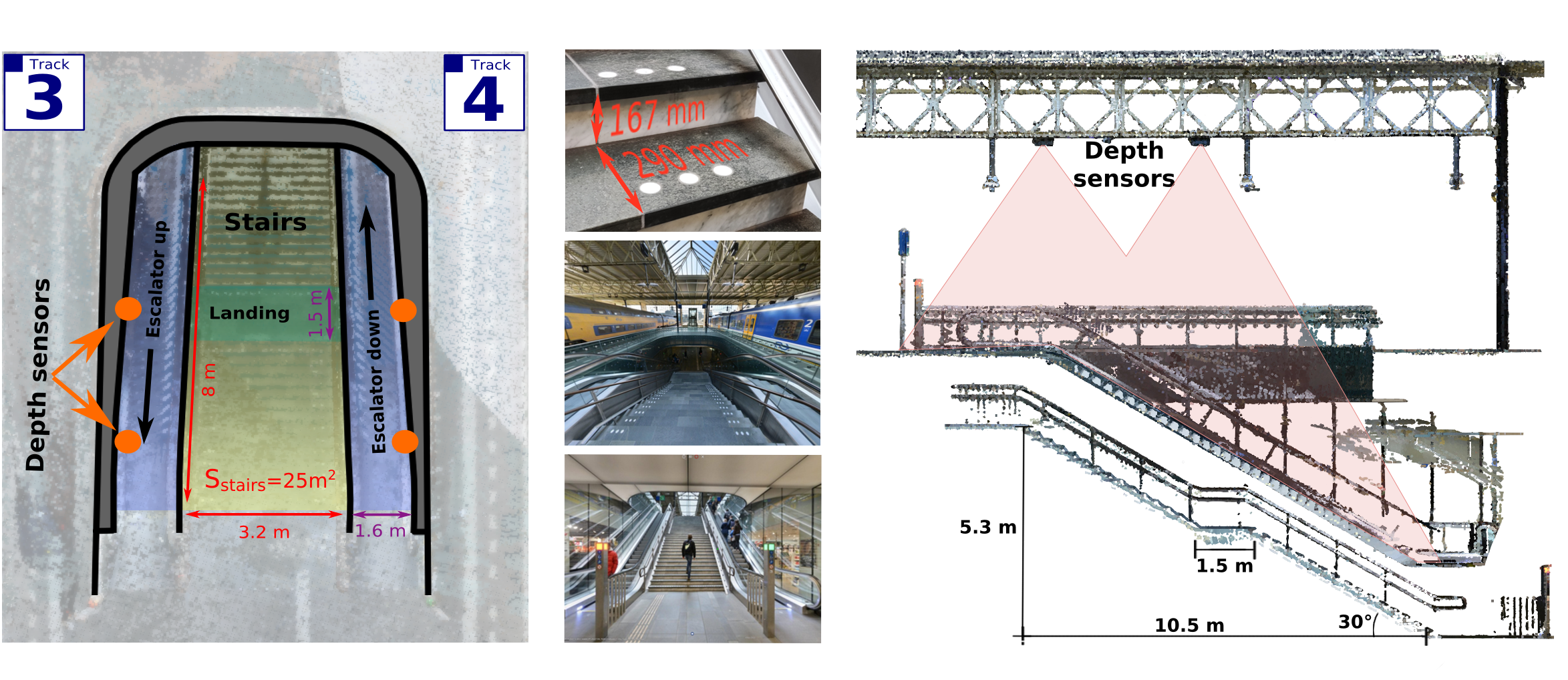}
    \put(-3.0, 38.5){(a)}
    \put(33.5, 38.5){(b)}
    \put(33.5, 26.5){(c)}
    \put(33.5, 13.0){(d)}
    \put(52.2, 38.5){(e)}
  \end{overpic}

  \caption{Schematic representation of the pedestrian measurement
    setup at Eindhoven Central Station (NL). (a) Top view of the
    staircase (yellow) and escalators (blue) from the perspective of
    the four depth sensors (orange dots). Escalator on the side of
    track 4 descending towards the tunnel, escalator on the side of
    track 3 ascending towards the platform. A staircase landing
    (green) is located halfway up the staircase. (b) Close-up picture
    of the staircase with two treads of the staircase. Each tread
    has rise $h=167\;$mm and run $r=290\;$mm (c,d) Pictures of the
    staircase-escalator system in (c) ascending direction, and (d)
    descending direction. (e) Vertical cross-section of the escalator
    and staircase. The depth sensors (black rectangles) are mounted to
    trusses on the ceiling of the train station. The sensor view cones
    are highlighted with a red color. The staircase and escalator both
    have a inclination of $\theta = 30^{\circ}$. For clarity, we
    applied in the image a small offset to the cross-section of the
    staircase.}\label{fig:setup}
\end{figure}

\section{Pedestrian Sensing at Large Scales: from depth maps to pedestrian trajectories
        }\label{sec:data-collection}

\subsection{Measurement setup at Eindhoven Train Station (NL)}\label{sec:experimental-setup}

The data presented in this work has been collected during a
1-year period between April 2021 and May 2022. 
Eindhoven Central Station (CS) is the fifth-largest train station 
in the Netherlands with 3 platforms and 6 train tracks. 
During this period of time, the station has been used on 
average by 46k travelers per day~\cite{NS-report}, a number dimmed
by the COVID pandemics as shown by the 77k daily users reported for the year 2019.
We have continuously recorded all the movements of pedestrians traversing a
highly trafficked staircase and escalator during the operational hours of
the train station (6 am -- 22 pm). The staircase-escalator system,
Fig.~\ref{fig:setup}, is the main entry point to train tracks 3 and
4. This means that its usage is highly correlated with the train
schedule. Typically, boarding passengers arrive at the platform
relatively scattered around the train arrival time. On the other hand,
alighting passengers reach the staircase in large compact groups as soon
as they leave their train.  

To perform our recordings, we have developed and
employed a custom system composed of a grid of four overhead depth
sensors. These sensors are controlled via dedicated software
performing depth recording, sensor fusion, pedestrian localization and
Lagrangian tracking. 
The measurement campaign allowed to acquire a dataset boasting about 3
million trajectories: an average of 10 thousand per day during
weekdays and 3 thousand per day during weekends.
The observed staircase has a relatively narrow width of
$w_{\textrm{stairs}}=3.2\;$m. This makes it a notorious bottleneck for
the pedestrian flow through the train station.
The stairs span a total elevation change of $h_{stairs}=5.3\;$m and a
horizontal distance of $l_{stairs}=10.5\;$m with a slope of
$\theta = 30^{\circ}$. The staircase consists of two flights of
stairs separated by a flat staircase landing halfway up the stairs,
and with length $l_{landing}=1.5\;$m.
The individual treads have standard dimensions (rise $h=167\;$mm, run
$r=290\;$mm), which are comparable to those from previous studies
present in the literature (cf. Table~\ref{tab:free-stream-vel}), 
making the dynamics on such staircase representative for a general case. 
The escalators are located on
either side of the staircase with a right-hand orientation.
They have standard escalator dimensions rise $h=220\;$mm, and run
$r=400\;$mm. The escalator has a slope of $30^{\circ}$ and a
horizontal velocity of $v = 0.6\;$m/s. 

\subsection{High-accuracy pedestrian tracking via depth-based 3D computer vision}\label{sec:computer-vision}

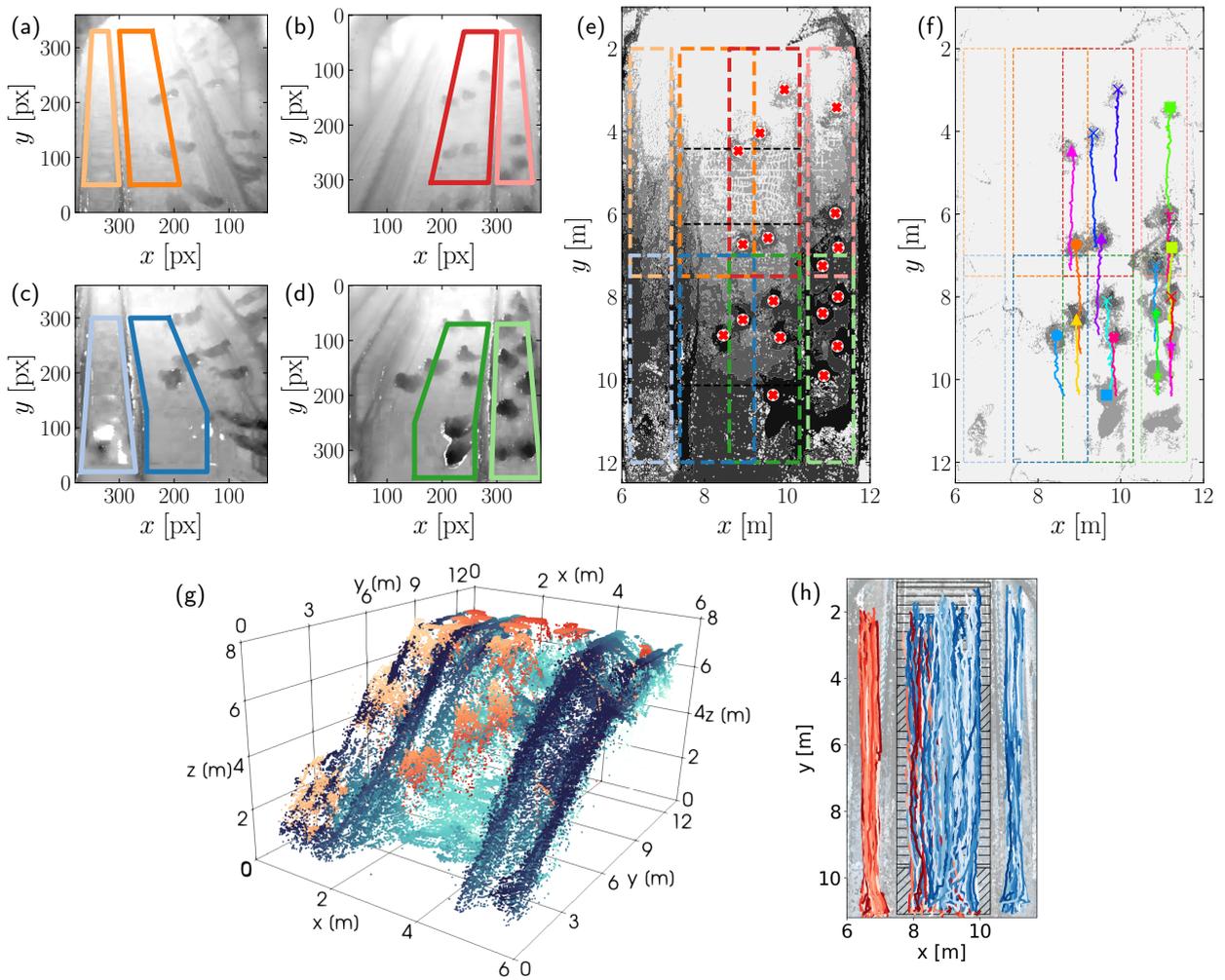
\begin{figure}[t]
  \centering 
  \begin{overpic}[width=\linewidth]{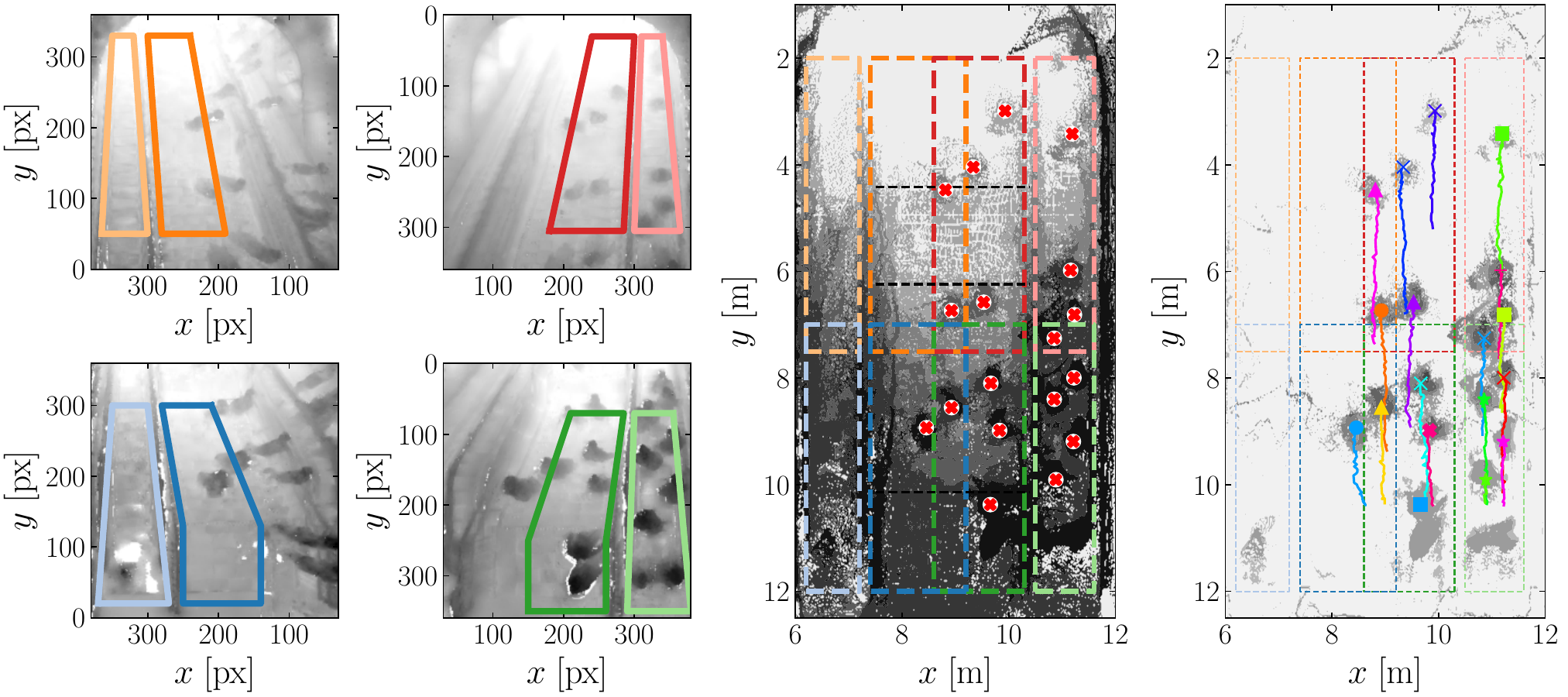}
    \put( 0.5, 42.0){(a)}
    \put(23.0, 42.0){(b)}
    \put( 0.5, 20.0){(c)}
    \put(23.0, 20.0){(d)}
    \put(47.0, 42.0){(e)}
    \put(75.0, 42.0){(f)}    
  \end{overpic}
  \begin{overpic}[width=.50\linewidth]{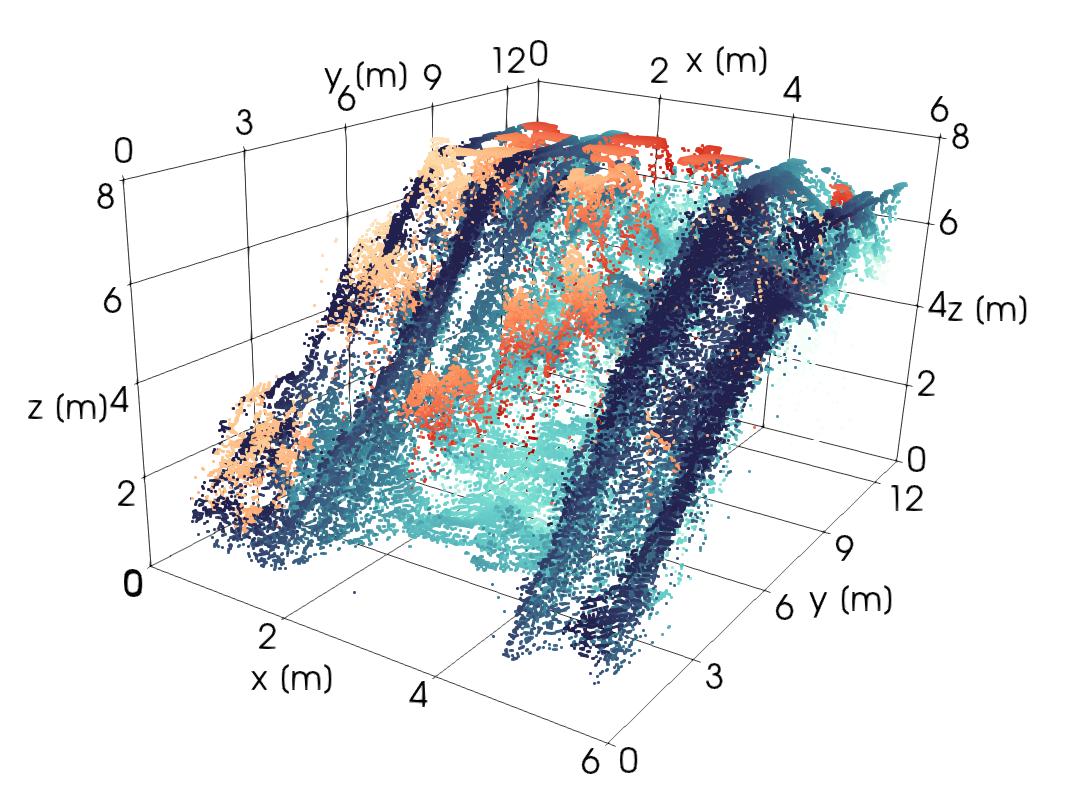}
    \put( 0.5, 63.0){(g)}
  \end{overpic}
  \begin{overpic}[width=.22\linewidth]{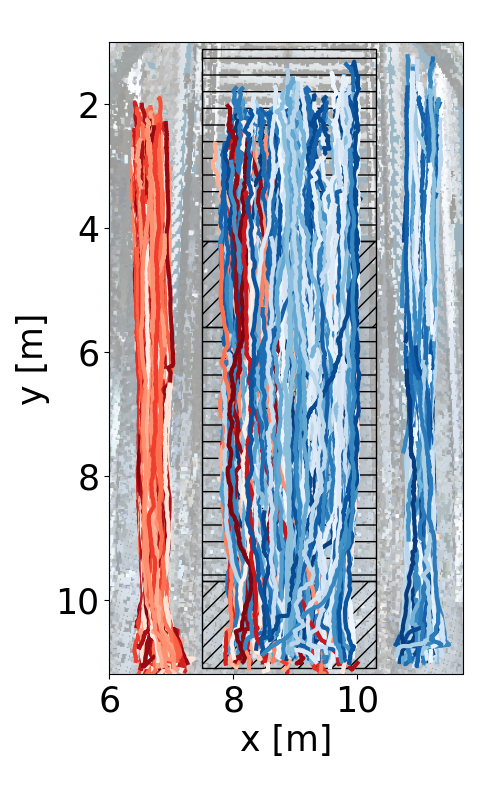}
    \put( 0.5, 90.0){(h)}
  \end{overpic}
  \caption{Depthmap images at different stages of the image
    processing. (a-d) Raw recordings of the four depth sensors. Each
    depth image is imposed with two regions of interest (ROI), one on
    the staircase (dark hue) and one on the escalator (light hue). (e,
    f) Depth images after perspective correction and sensor
    merging. From the ROIs we observe the overlapping parts of the
    sensor domains. We report for the pedestrian localizations in this
    frame together with the Lagrangian trajectories over the previous
    $10$ seconds. (e) Without background subtraction and (f) with
    background subtraction. (g) 3D pointcloud of the perspective
    corrected and stitched depth images. The background is colored
    with blue tones and the foreground is colored with red tones. (h)
    Recorded trajectories across the staircase. Ascending trajectories
    are colored with a red hue and descending trajectories with a blue
    hue.}\label{fig:depthmap}
\end{figure}

We summarize here technical and algorithmic aspects of our pedestrian
tracking system. We acquire raw data through a grid of overhead depth
sensors recording at 30 frames per second. We employ Stereolabs Zed 2
sensors~\cite{zed2}, driven by Nvidia Jetson TX2
GPUs~\cite{jetson}. The sensors are mounted on horizontal trusses
above the staircase-escalator system as presented in
Fig.~\ref{fig:setup}. Depth sensors measure the so-called depth
(\textit{or distance}) field, i.e.\ they probe the distance between
each observed point and the sensor plane. This signal is typically
represented and memorized in terms of gray scale images which, in our
case, have resolution $640\times 360\;$px$^2$
(cf. Fig~\ref{fig:depthmap}). The gray level encodes distance with
shades that become brighter (Fig.~\ref{fig:depthmap}a-f) as distance
increases. Zed sensors measure depth through a stereo vision approach:
they acquire two simultaneous color views of the same scene from CCD
sensors having a lateral offset of $12\;$cm. Similarly to the human
brain, the two images are used to estimate distance.  This distance
reconstruction, generally referred to as stereo matching, is performed
in real-time through the Zed API running on the local Nvidia Jetson
TX2 card. We design our system in such a way that no color image
(possibly privacy infringing) is explicitly downloaded from the
sensors.
We position the sensors in such a way that their view cone is
partially overlapping. This configuration enables
continuous coverage of the whole staircase area and of a volume that
spans from the stairs up to the ground level plane. 

Streams of depth maps are at the base of our trajectory measurement
approach. 
The processing pipeline, applied to approximately $400\;$GB of depth images per day,
is summarized in Fig.~\ref{fig:pipeline} and
involves six main stages. Some of these stages have been described in
previous works by the same authors~\cite{corbetta-pre-2017,corbetta-pre-2018,willems-sr-2020}. 
For the sake of completeness we report in what follows on all stages. 

We can formalize a depth image with the field $(\xv, d(\xv))$,
which maps each pixel, $\xv$, in the image with its distance $d(\xv)$ from the camera.
Streams of depth images undergo the following six stages:
\begin{figure}[t]
  \centering
  \begin{overpic}[width=\linewidth, trim = 250 50 210 20,clip]{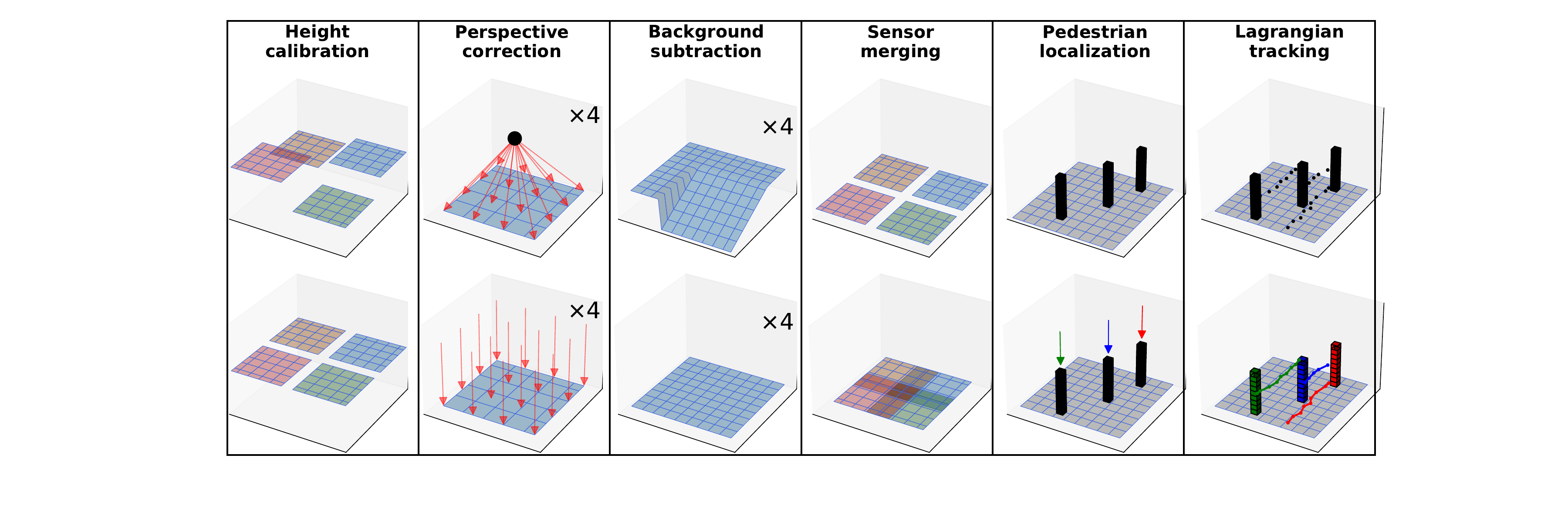}
    \put( 0.5, 37.3){(a)}
    \put(17.0, 37.3){(b)}
    \put(33.6, 37.3){(c)}
    \put(50.2, 37.3){(d)}
    \put(66.8, 37.3){(e)}
    \put(83.5, 37.3){(f)}
  \end{overpic}
  \caption{ Synthetic examples of the processing
    steps for pedestrian tracking on inclined planes. 
    (a) Height calibration, establishing agreement on the depth measurements. 
    (b) Perspective correction. (c) Background subtraction. 
    The foreground will reveal all (new) objects removing all 
    terrain/ flattening the ground. 
    (d) Merging all sensor images thereby creating one shared 
    coordinate system. (e) Pedestrian localization. 
    (f) Lagrangian time tracking.
    }\label{fig:pipeline}
\end{figure}
\begin{enumerate}
\item \textbf{Height calibration} We observe that depth measurements
  present noise that affects the depth field uniformly and in
  a multiplicative way. In other terms, we measure the field
  $Ad(\xv)$, where $A$ is a constant changing in time in dependence on
  the global scene illumination. We use a fixed spatial reference
  (depth of the platform) to normalize away the multiplicative
  constant $A$ and obtain $d(\xv)$. This correction appears to be
  necessary with Zed sensors. Other sensors used by the same authors,
  e.g., Microsoft Kinects, did not need it.
\item \textbf{Perspective correction} We perform an axonometric
  transformation of the depth image, conceptually equivalent to
  bringing the sensor to an infinite altitude. This turns the view
  from conical to cylindrical and renders all the sight rays vertical
  and mutually parallel. For an ideal pinhole view, and up to a
  multiplicative constant $C$, this transformation re-projects a
  point $\xv$ in image space as
    \begin{equation}\label{eq:reprojection}
      (\xv, d(\xv)) \rightarrow  ( C \xv d(\xv),  d(\xv)).
    \end{equation}
    This transformation would ensure that points that are vertically
    aligned in space get mapped to the same location $C \xv
    d(\xv)$. 

    We observed, however, that the ideal model in
    Eq.~\eqref{eq:reprojection} is still insufficient for a precise
    mapping of the space. This holds especially when the distance to
    the sensor (depth) changes significantly (i.e.\ moving across the
    stairs). To compensate on this, we consider a second order correction model:
    \begin{equation}\label{eq:second-order-projection}
      (\xv, d(\xv)) \rightarrow (\xv, c_1\cdot d(\xv)^2 + c_2\cdot d(\xv) + c_3),
    \end{equation}
    with $c_1$, $c_2$ and $c_3$ free parameters. We estimated the parameters by 
    moving calibration targets on the scene (cardboard boxes of known size), 
    and ensuring that the railings of the escalator, once projected 
    with Eq.~\eqref{eq:second-order-projection}, remain mutually parallel.

    We generate an axonometric depth image retaining for each
    vertically aligned point the one closest to the sensor.
  \item \textbf{Background subtraction} To reduce localization
    artifacts, we retain the foreground of each depth signal. That is,
    only the pixels that are sufficiently closer to the sensor plane
    than a background estimated in absence of crowding.  This
    transformation operates as follows:
    \begin{equation}\label{eq:background}
      (\xv, d(\xv)) \rightarrow (\xv, d(\xv) - \min_{\Delta t} d(\xv)),
    \end{equation}
    with $\min_{\Delta t} d(\tilde{\xv})$ the minimum depth value in
  the image over a small time window $\Delta t$. Typically, this time 
  window is chosen in very dilute conditions to ensure the absence of crowding. 
  In our case the background was recorded at night when the facility is closed.
\item \textbf{Depth data fusion} Having defined an 
  axonometric view for the four depth streams, it is then possible 
  to juxtapose them to generate a single depth
  image covering the entire area. In the overlapping regions, pixels
  with depth closest to the camera are retained. The merging is
  performed such to ensure that the reference cardboard boxes preserve
  their shape and area throughout the entire fused depth image.
\item \textbf{Localization} We perform localization using the YOLOv7
  localization algorithm~\cite{yolov7-arxiv-2022}.  YOLO (You Only
  Look Once) has recently emerged as a widespread adopted algorithm
  for real-time object detection. Unlike traditional methods that
  employ multi-stage pipelines, YOLO takes a unified approach by
  performing object detection in a single pass.
  In short, a given input image gets divided into a grid and the
  algorithm simultaneously make predictions for bounding boxes and
  class probabilities for each grid cell by leveraging deep neural
  networks (NNs).
  We have trained a NN from scratch by hand-annotating depth images in
  an active learning fashion. In other terms, after a first annotation
  session (about 15 frames), we proceeded correcting the output of
  the localization to improve the training dataset.
  This has allowed for high quality tracking, outperforming previous
  approaches (\cite{pouw-cd-2021}), specially in highly dense
  conditions.
  In Table~\ref{tab:yolo}, we report the performance achieved by the
  YOLO algorithm using as test data two hours of hand-annotated depth
  images.  We report precision, recall, and F1-score for different
  values of the local density. We refer the reader to
  Appendix~\ref{app:yolo} for further details.
\item \textbf{Tracking} We perform Lagrangian tracking of bounding box
  centroid via the Trackpy library \cite{trackpy}. We cross-validate the obtained
  trajectories against the optical flow aiming at detecting false
  positives and tracking errors.
\end{enumerate}
\begin{table}[htb]
  \caption{Performance of our YOLO-based pedestrian localization algorithm against about 1000 hand-annotated frames. We report standard quality metrics: precision, recall, and F1 score. For details about the algorithm and its validation, and formal definitions of the performance metrics, see Appendix~\ref{app:yolo}.
  }\label{tab:yolo}
  \begin{tabular}{ccccccc} 
      \toprule
      {Density Range ($\rm{ped/m^2}$)}  & {True positive} & {False positive} & {False negative} &  {Precision} & {Recall} & {F1 Score} \\ 
      \midrule
      0.55 - 0.75 & 373   &   0 &    2 & 1.0    & 0.9947 & 0.9973 \\
      0.75 - 0.95 & 1301  &   6 &   18 & 0.9954 & 0.9864 & 0.9909 \\
      0.95 - 1.15 & 2012  &  13 &   80 & 0.9936 & 0.9618 & 0.9774 \\
      1.15 - 1.35 & 2735  &  27 &  174 & 0.9902 & 0.9402 & 0.9646 \\
      1.35 - 1.55 & 3140  &  20 &  197 & 0.9937 & 0.9410 & 0.9666 \\
    \bottomrule
  \end{tabular}
\end{table}

\begin{figure}[t]
  \centering
  \begin{overpic}[width=\linewidth]{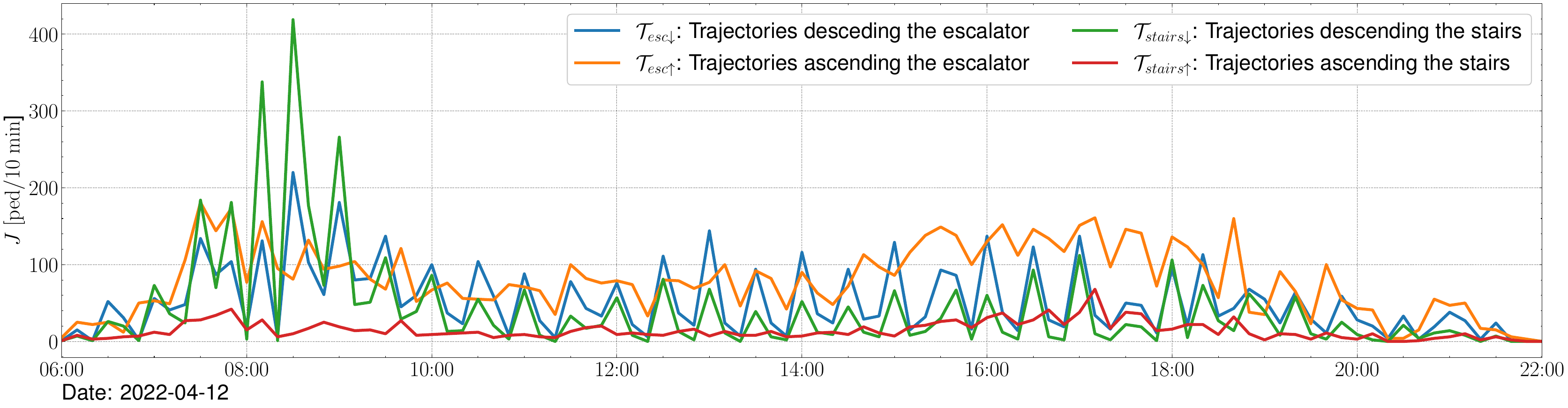} 
    \put( 0.5, 21.0){(a)}
  \end{overpic}  
  \begin{overpic}[width=.333\linewidth]{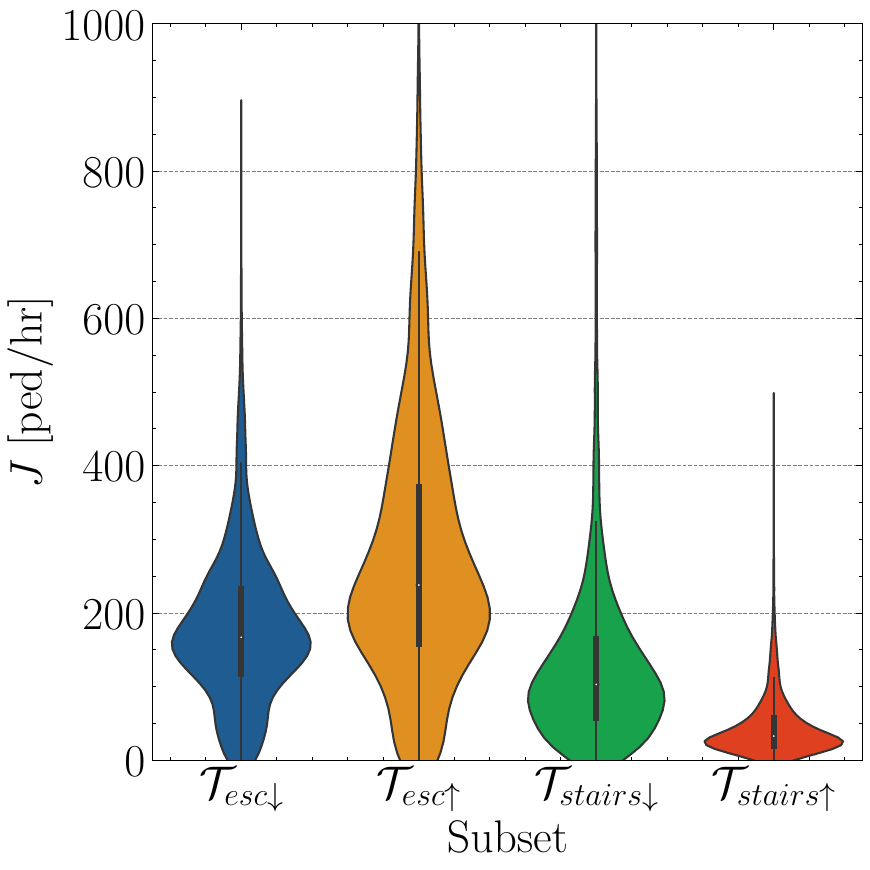} 
    \put( 0.5, 90.0){(b)}
  \end{overpic}  
  \begin{overpic}[width=.32\linewidth]{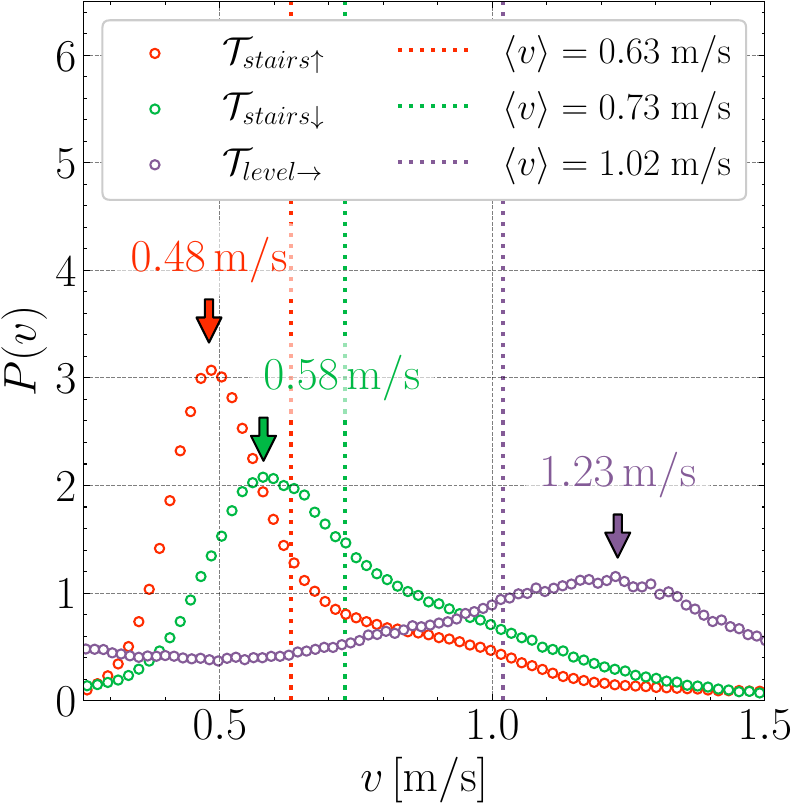} 
    \put( -1.0, 93.0){(c)}
  \end{overpic}  
  \begin{overpic}[width=.333\linewidth]{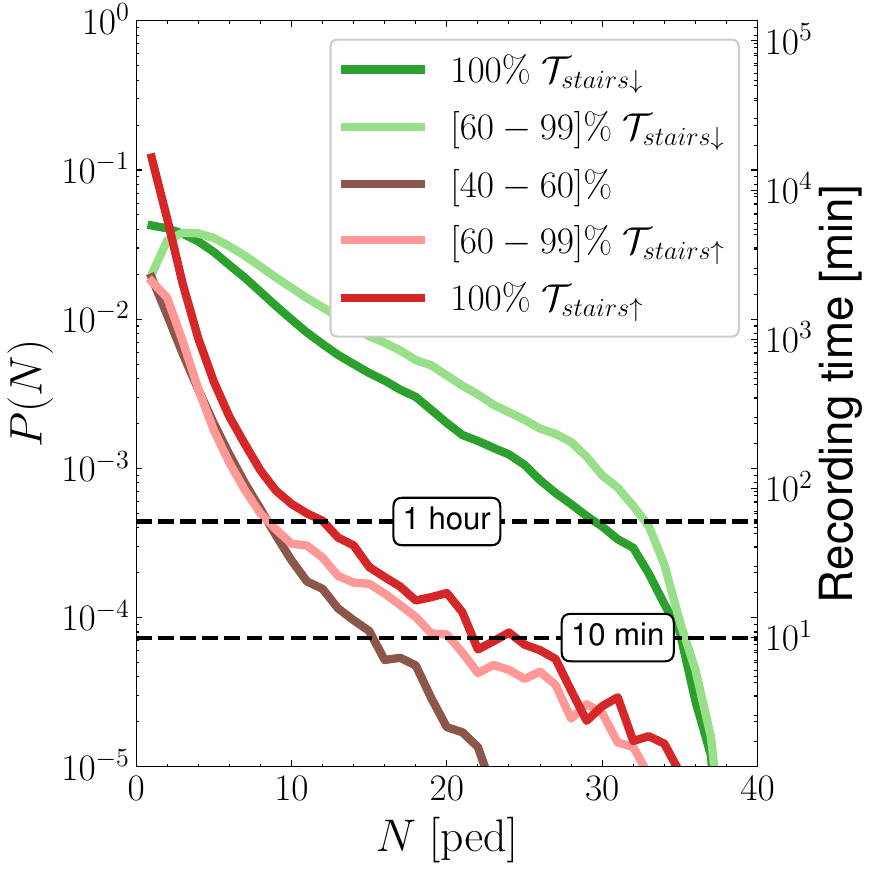}
    \put( 0.5, 90.0){(d)}
  \end{overpic}        
  \caption{(a) Pedestrian flux, $J$, across the staircase ascending
    $\mathcal{T}_{stairs\uparrow}$ (red) and descending
    $\mathcal{T}_{stairs\downarrow}$ (green) or using the escalators
    to ascend $\mathcal{T}_{esc\uparrow}$ (yellow) and descend
    $\mathcal{T}_{esc\uparrow}$ (blue) per 10-minute time window for a
    typical working day. We observe a morning rush hour at 08:00-09:00
    and afternoon rush hour 16:00-18:00. The morning is dominated by
    people going down to the tunnel using the stairs (green) or the
    escalator (blue), and the afternoon by pedestrians in the other
    direction going towards the platform using the escalator
    (orange). (b) Boxplots of the pedestrian flux, $J$, per movement
    mode in pedestrians per hour. (c) Probability distribution of the
    free-flow walking velocity on the staircase ascending (green) and
    descending (red). For comparison we also report the free-flow
    walking velocity on level ground, $\mathcal{T}_{level\rightarrow}$
    (purple). We annotate with arrows the most probable walking
    velocities i.e.\ $v_{stairs\uparrow} = 0.48$ m/s,
    $v_{stairs\downarrow} = 0.58$ m/s and
    $v_{level\rightarrow} = 1.23$ m/s. (d) Probability distribution of
    the flow dynamics as a function of the stairs occupation
    $N$. We report five possible flow dynamics: (1)
    unidirectional flows descending
    $100\%\;\mathcal{T}_{stairs\downarrow}$ (green) and (2) ascending
    $100\%\;\mathcal{T}_{stairs\uparrow}$ (red) (3) unbalanced
    bidirectional flows dominated by descending pedestrians
    $[60-99]\%\;\mathcal{T}_{stairs\downarrow}$ (light green) and (4)
    dominated by ascending $[60-99]\%\;\mathcal{T}_{stairs\uparrow}$
    (light red), and (5) balanced bidirectional flow (brown) with a
    $[40-60]\%$ ratio between ascending and descending. We observe on
    the staircase mainly unidirectional and unbalanced bidirectional
    flows in the descending direction. We employ a second y axis on the right side of the figure with the recording times in minutes to indicate the size of the data for each of the five possible flow dynamics. Additionally, we impose two horizontal lines signalling thresholds for 10 minutes and 1 hour of recordings respectively. For descending motions we have significant recordings ($>10$ min) for pedestrian occupancy up to $N<35$. For descending motion however we have few recordings for values of $N>20$ pedestrians.}\label{fig:stats}
\end{figure}

\begin{figure}[t]
  \centering
  \begin{overpic}[width=.8\linewidth]{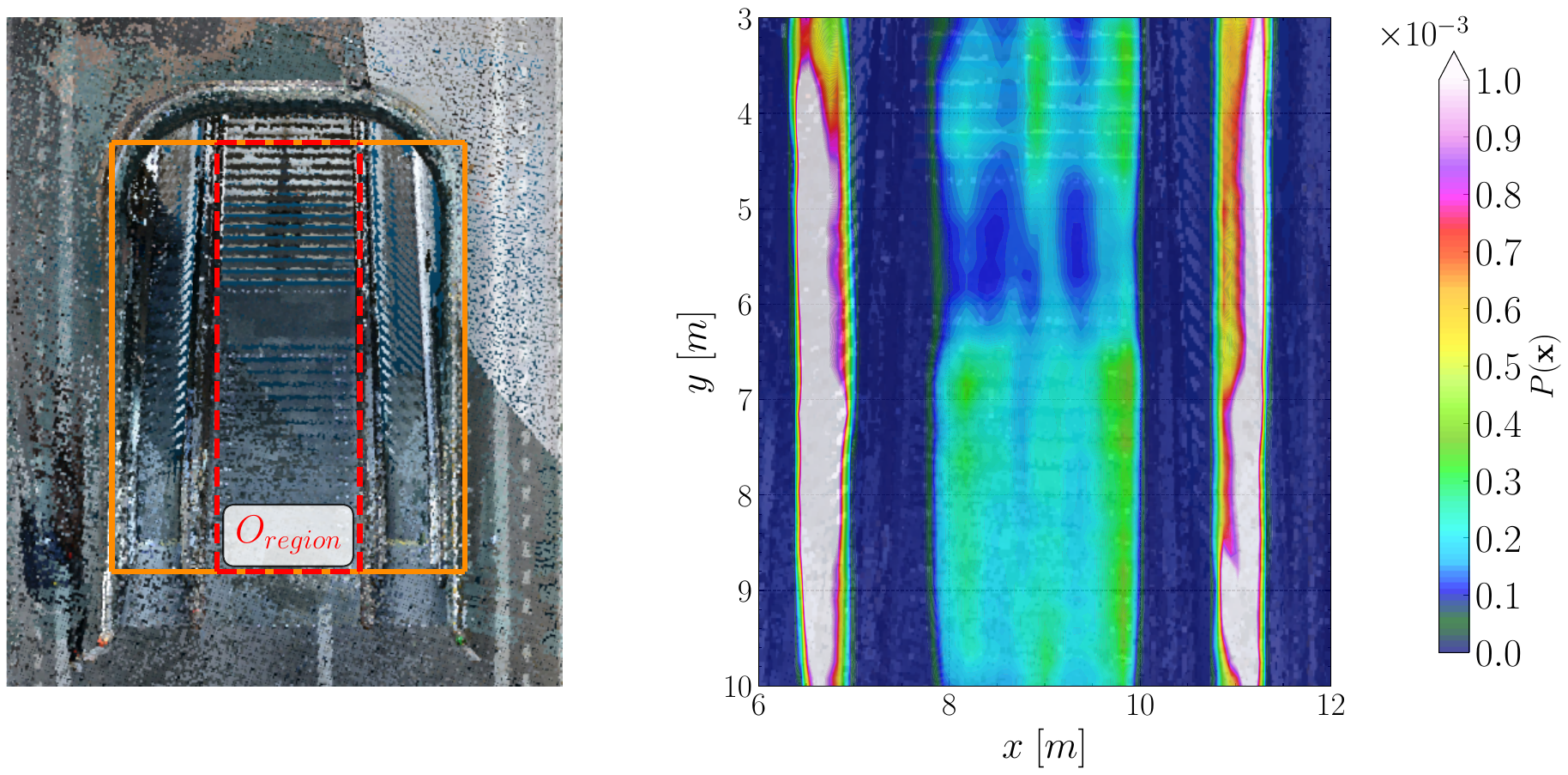}
    \put(-5.5, 43.0){(a)}
    \put(40.0, 43.0){(b)}
  \end{overpic}  
  \begin{overpic}[width=\linewidth]{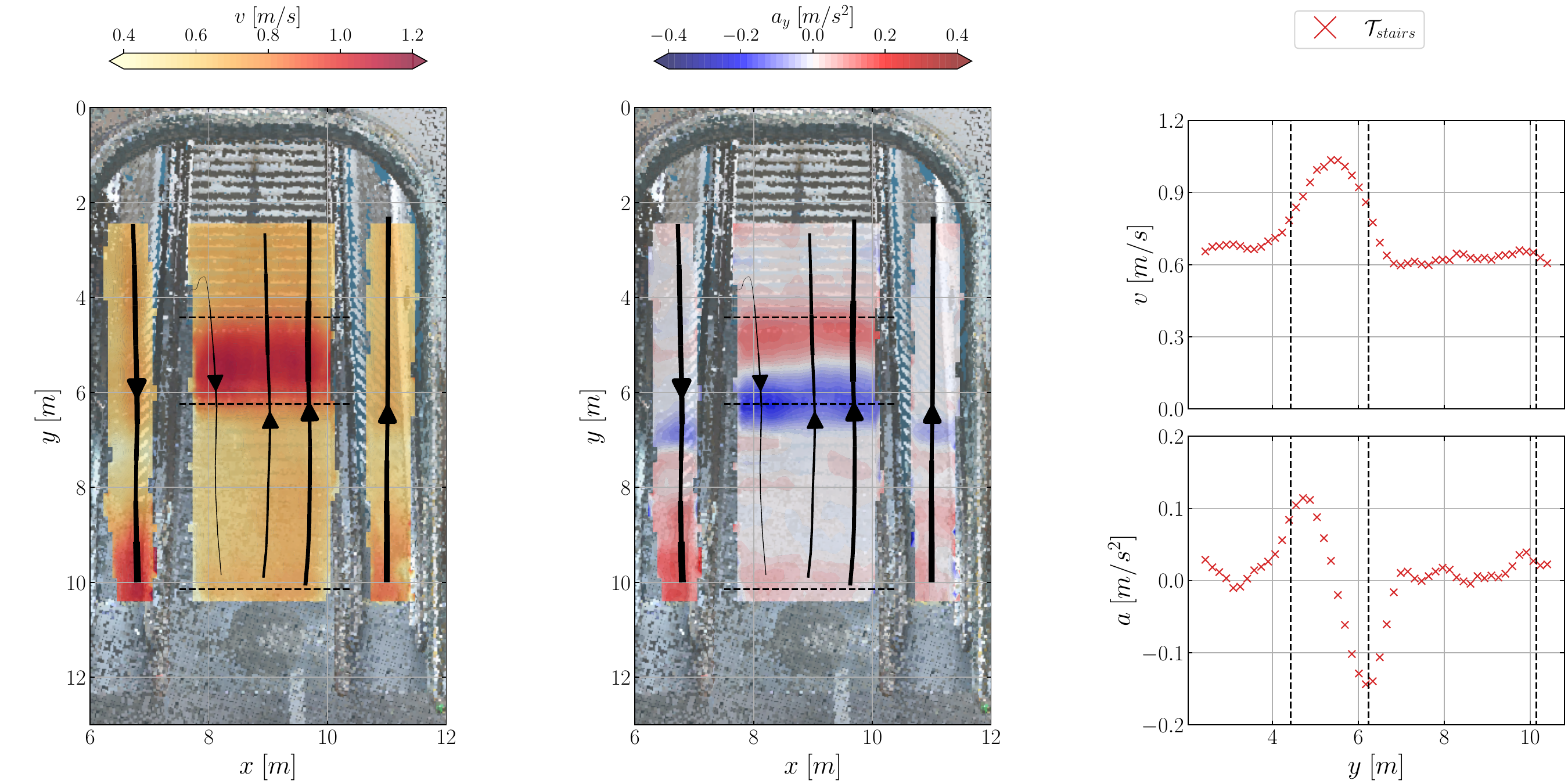}
    \put( 0.5, 41.0){(c)}
    \put(35.0, 41.0){(d)}
    \put(70.0, 41.0){(e)}
    \put(70.0, 21.0){(f)}
  \end{overpic}    
  \caption{(a) Overhead image of the staircase considered. Our sensors covered area within the orange rectangle. We consider here the staircase region $O_{region}$ bounded by the dashed red line.  (b) Probability distribution function of pedestrian position represented as a heatmap. We observe three
    walking lanes across the staircase. (c, d) The distributions of
    velocity and acceleration across the staircase-escalator
    system. We observe a higher walking velocity on the flat staircase
    landing. Both images are imposed with flow vectors indicating the
    direction of the flow. We observe that pedestrians walk on the
    right side of the staircase following the direction of the
    escalators. (e, f) Cross-sections of the velocity (e) and
    acceleration (f) on the staircase. All distributions are computed by considering the entire dataset $\mathcal{T}$.
  }\label{fig:heatmaps}
\end{figure}
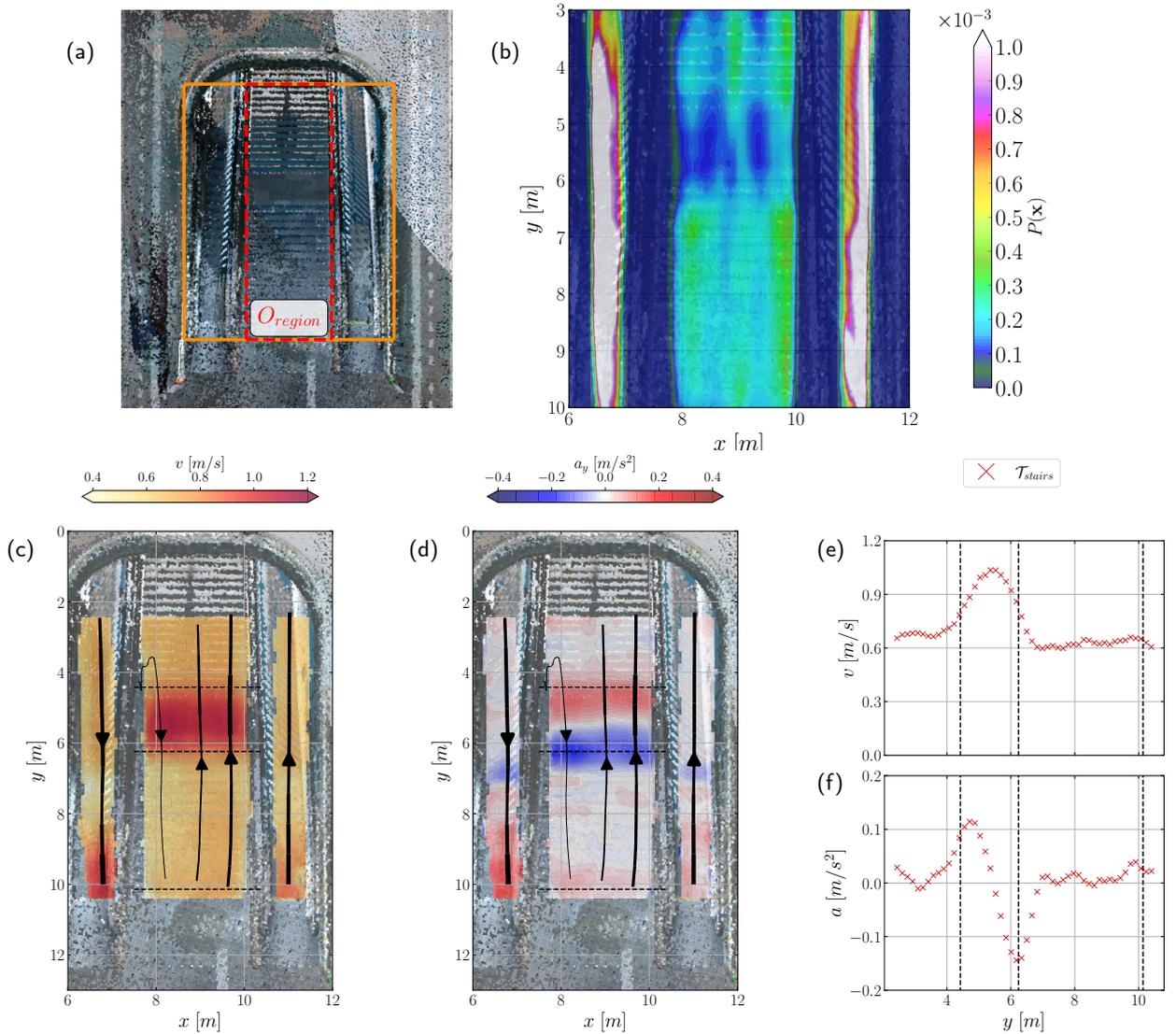

\subsection{Pedestrian velocity and density: operational definitions}\label{sec:density-vel}
In this section, we define the notation and a few useful quantities, akin to those used by \cite{saberi-trr-2014, hoogendoorn-trc-2018}, to guide the reader through the analysis presented in Section 3. In what follows, we will refer to the trajectory dataset, obtained following the procedure described in the previous section, with the symbol $\trajds$.

We consider a coordinate system $\xv = (x,y)$ such that the $x$ and
$y$ axes are, respectively, transversal and longitudinal with respect
to the ascending/descending direction (cf. Fig.~\ref{fig:depthmap}).
In other terms, we consider a projection of the staircase on the
horizontal plane. We denote with $\vv(t) = ( v_x(t), v_y(t) )$ the
instantaneous velocity of a pedestrian, which we compute applying
Savitzky-Golay filtering~\cite{savitzky-ac-1964} on trajectory data,
and likewise for the instantaneous acceleration
$\textbf{a}(t) = ( a_x(t), a_y(t) ) $. Additionally, we define the frame-average walking velocity, $v_{t,\mathcal{T}}$, at frame, $t$, for a trajectory subset,
$\mathcal{T}$:
\begin{equation}\label{eq:walking-velocity}
  v_{t,\mathcal{T}}
  =
  \frac{1}{N_{t,\mathcal{T}}} \sum^{N_{t,\mathcal{T}}}_{i=1}v_{i} ,
\end{equation}
with $v_{i}$ the velocity of pedestrian $i$ and $N_{t,\mathcal{T}}$
the number of pedestrians in frame $t$ part of subset $\mathcal{T}$. 
Due to the subset being almost obviously deductible from the context, 
for ease of notation we shall indicate the average velocity simply with $v$.

We will distinguish trajectories in our dataset, $\trajds$, based on two criteria:
\begin{itemize}
\item Direction, either downstairs or upstairs. We determine the
  direction by setting a threshold on the average longitudinal
  velocity of each trajectory, $\avgg{v_y}$. We indicate respectively
  with $\trajdsD$ and $\trajdsU$ the sets of trajectories in
  downstairs and upstairs direction. In formulas, it holds
    \begin{equation}
      \trajdsD
      = 
      \left\{\gamma \in \trajds\; | \; \avgg{v_{y}}  < - V_t  \right\} 
      \qquad  \qquad 
      \trajdsU  
      =
      \left\{\gamma \in \trajds\; | \; \avgg{v_{y}}  >  V_t  \right\},
    \end{equation}
    with the threshold $V_t = 0.2\;$m/s.
  \item Movement mode, either via the stairs or via the escalators. 
    We distinguish between the movement modes based on the position. We
    define the subsets $\mathcal{T}_{\rm esc}$ and $\mathcal{T}_{\rm stairs}$ as
    the sets consisting of trajectories with all coordinate observations
    located either on the escalators or on the staircase area, respectively. In
    formulas, these read
    \begin{equation}
      \mathcal{T}_{\rm esc} 
      = 
      \left\{\gamma \in \trajds\; | \;\textbf{x} \in S_{\rm esc}\right\} 
      \qquad  \qquad 
      \mathcal{T}_{\rm stairs} 
      = 
      \left\{\gamma \in \trajds \; | \;\textbf{x} \in S_{\rm stairs} \right\}.
    \end{equation}
\end{itemize}
These criteria yield four disjoint subsets
\begin{itemize}
  \item $\trajds_{stairs \uparrow} = \trajds_{stairs} \cap \trajdsU$,
    trajectories going in upstairs direction via the staircase;
  \item $\mathcal{T}_{stairs\downarrow} = \trajds_{stairs} \cap \trajdsD$, 
    trajectories going in downstairs direction via the staircase;
  \item $\mathcal{T}_{esc\uparrow} = \trajds_{esc} \cap \trajdsU$, 
    trajectories going in upstairs direction via the escalators;
  \item $\mathcal{T}_{esc\downarrow} = \trajds_{esc} \cap \trajdsD$,  
    trajectories going in downstairs direction via the escalators;
\end{itemize}
Note that this categorization excludes rare cases such as pedestrians
inverting their trajectories, which will be neglected from our analysis.

We conclude this section discussing the frame by frame estimation of 
the density, which is crucial for the computation of the fundamental diagrams. Within the literature on pedestrian dynamics, numerous methods have been introduced to estimate pedestrian density. For a comprehensive comparison of these methods, we refer to \cite{steffen-physa-2010, zhang-josm-2011,duives-physa-2015}. Traditionally, the pedestrian density $\rho$ has been computed using
the hydrodynamic definition: 
\begin{equation}\label{eq:classical-density}
  \rho = \frac{N}{\bar S},
\end{equation}
where $N$ is the number of pedestrians in the observation
area $\bar S$ (e.g.,~\cite{steffen-physa-2010,
fruin-book-1971}). This definition provides robust estimates in case
of spatially uniform flows over surfaces much larger than the
pedestrian diameter (i.e.\ as the continuum limit approximation becomes
accurate).  On the other hand, this definition strongly depends on the
size of the measurement area, $\bar S$,~\cite{zhang-josm-2011},
while averaging out local density fluctuations.
These issues have been circumvented
considering a Voronoi tessellation centered in the pedestrians
position (e.g.~\cite{steffen-physa-2010}),
enabling a local definition of the density field as the reciprocal of
the local Voronoi cell area. This method provides robust, although 
computationally expensive, density estimates in the bulk of the flow 
and at sufficiently high density levels. 
On the other hand, in the absence of neighboring pedestrians, the Voronoi 
cells result unbounded with local density degenerating to zero. 
This issue can be mitigated by limiting the size of the cells 
to the boundaries of the observed geometry (hence all cells result closed) 
and/or by thresholding each cell (to prevent density underestimation).

In order to deal efficiently with high statistics data and prevent unbounded
cells, we consider here an approach that builds on bounding individual
areas from the start, echoing the concept of personal space. The personal space is defined as the space a walking pedestrian
tries to maintain around the body. While it is understood to be
elliptical (axes: $2\;$m in longitudinal and $0.5\;$m in lateral
direction according to~\cite{gerinlajoie-gp-2008}),
we approximate the individual personal space of the pedestrian, $i$,
as a circular region with radius $R = 0.75\;$m and area $\pi R^{2} = 1.86\;$m$^{2}$.
In level-of-service terms, this is at the interface between level A
(free-flow) and B (slightly restricted flow), 
as defined by~\cite{fruin-book-1971}.
We compute the instantaneous density as the ratio between the number
of observed pedestrians and the union of
all the personal spaces,
\begin{equation}\label{eq:area}
  \hat S_{stairs} = \mbox{Area}\left(\bigcup^{N}_{i=1} S_{i} \cap O_{region} \right)\  \leq \  \mbox{Area}(O_{region}) = \bar{S} \approx 25~\rm{m^2},
\end{equation}
i.e.\ 
\begin{equation}\label{eq:density-hat}
  \hat\rho = \frac{N}{\hat S_{stairs}},
\end{equation}
where $O_{region}$ indicates the observation region, that is used as outer limit of the personal areas. 
Note that according to Eq.~\eqref{eq:density-hat}, the density has the following lower and upper bounds:
\begin{equation}\label{eq:density-bounds}
  \max\left\{ \frac{1}{\pi \cdot R^2} , \frac{N}{\bar S} \right\} \leq \hat \rho \leq \frac{N}{\pi \cdot R^2}.
\end{equation}
The lower bound entails the maximum between two components. The
constant $1/(\pi R^2)=0.56\;$m$^{2}$ is attained when personal spaces
do not overlap. Moreover, due to $\hat S_{stairs}\leq \bar S$
(Eq.~(\ref{eq:area})), our density definition is always larger than
the hydrodynamic definition (Eq.~\eqref{eq:classical-density}, that for small $N$ can get arbitrarily
close to zero; see also Appendix~\ref{app:comparison}). The density
value $\hat \rho$ increases as personal spaces overlap (mutual
distances $d < 2\cdot R$), reducing the union in
Eq.~\eqref{eq:area}. This yields a theoretical upper bound
corresponding to the case in which all the personal areas perfectly
overlap. This is clearly a non-physical upper bound due to
necessary volume exclusions. More realistically, for the case of $N$
very large, personal areas will fill the available surface
$O_{region}$.  This assumption enables to provide a more realistic
asymptotic behavior to $\hat\rho$, which approaches the hydrodynamic
density (see also trends in Fig.~\ref{fig:area-vs-density}(b)).

\section{Results}\label{sec:results}

\subsection{Average dynamics: pedestrian flux and free-stream velocity}\label{sec:data-overview}

We start our analysis by providing an overview of the overall 
usage of the stairs-escalator system in terms of typical daily
fluxes, fluxes distribution, directionality and velocities.

\paragraph{Flux and flow partition.}
In Fig.~\ref{fig:stats}a we provide an example of pedestrian fluxes
recorded during a typical working day (2022-04-12), for the four
different subset of trajectories ($\trajds_{stairs \uparrow}$,
$\mathcal{T}_{stairs\downarrow}$, $\mathcal{T}_{esc\uparrow}$,
$\mathcal{T}_{esc\downarrow}$) previously defined in
Sec.~\ref{sec:data-collection}.  We observe high fluxes during rush
hours, with average peaks of $50~\rm{ped/m}$ in the morning (7 am - 9
am) where the majority of commuters descend from the platform to the
train station (blue and green curves), and in the afternoon (16 pm -
18 pm) where flows in the opposite direction (orange and red curves)
play a more relevant role, with average peaks of $20~\rm{ped/m}$. We
shall remark that the asymmetry between morning and afternoon is
specific to the train platform considered in our study, and is due to
the train schedule.

In Fig.~\ref{fig:stats}b we show aggregated flux statistics, in terms
of pedestrians per hour, for each one of the trajectory subsets.  The
data highlights a strong preference towards the usage of the
escalator. The average flux for $\mathcal{T}_{esc\uparrow}$
($\mathcal{T}_{esc\downarrow}$) is $276~\rm{ped/h}$
($180~\rm{ped/h}$), while for the staircase we report $46~\rm{ped/h}$
and $131~\rm{ped/h}$, respectively for $\trajds_{stairs \uparrow}$ and
$\mathcal{T}_{stairs\downarrow}$. The overall fraction of people
choosing the escalator over the staircase is $0.72$, which becomes
$0.86$ when considering only the subset of people going upstairs, and
$0.58$ when considering only people going downstairs.

While the staircase is prominently used in downstairs direction,
bidirectional flows also occur. In Fig.~\ref{fig:stats}d we report the
probability distribution of the flow compositions as a function of the
pedestrian occupation on the stairs. We consider 5 categories with
different direction ratios, respectively: 1,2) unidirectional flow
i.e.\ pedestrians all move downstairs
$100\% \;\mathcal{T}_{stairs\downarrow}$ or all pedestrians move
upstairs $100\% \;\mathcal{T}_{stairs\uparrow}$; 3) balanced
bidirectional $[40-60]\% \;\mathcal{T}_{stairs\uparrow\downarrow}$,
i.e.\ an almost equal number of people are simultaneously going
upstairs and downstairs; 4,5) unbalanced bidirectional flow, i.e.\ the
majority of pedestrians is going downstairs
$[60-99]\% \;\mathcal{T}_{stairs\downarrow}$ or the majority of
pedestrians is going upstairs
$[60-99]\% \;\mathcal{T}_{stairs\uparrow}$. We rarely observe a
balanced bidirectional flow, especially for high crowd densities. Considering the primarily unidirectional flow, the aforementioned asymmetry between morning and afternoon does not influence our study.

\paragraph{Free-stream velocity.}
\begin{figure}[t]
  \centering
  \begin{overpic}[width=\linewidth]{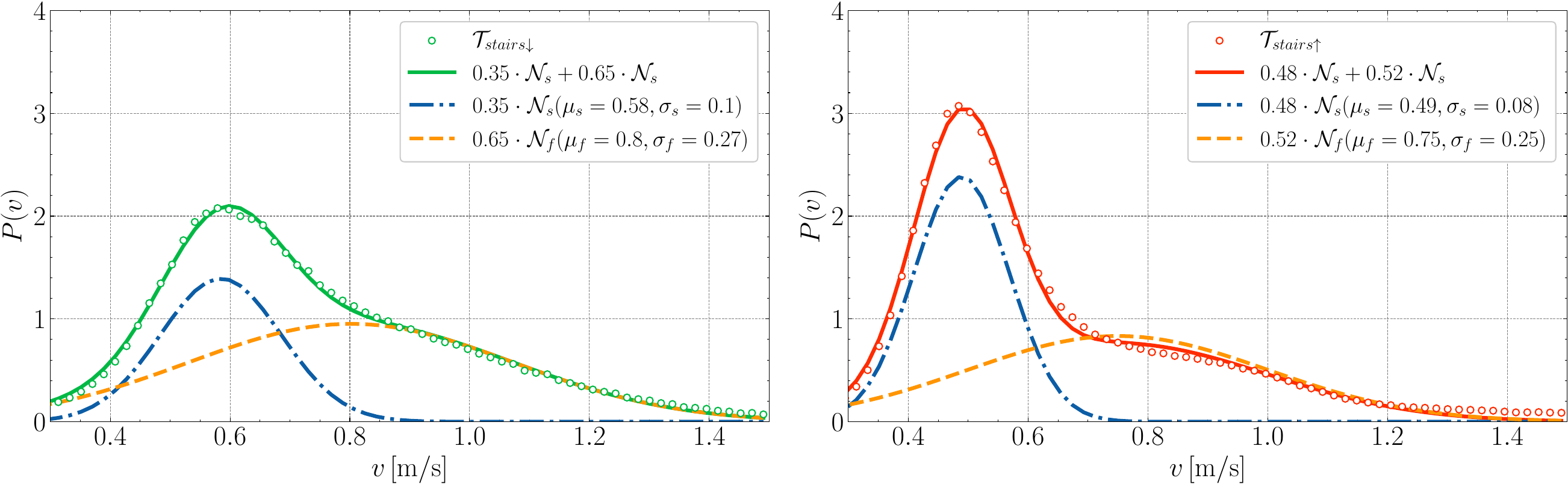}
    \put(-1.5, 30.0){(a)}
    \put(50.0, 30.0){(b)}
  \end{overpic}  
 
  \caption{Probability distribution of the free-stream walking
    velocity for (a) downstairs and (b) upstairs motion. We fit a
    mixture of two Gaussian distributions
    (Eq.~\eqref{eq:gauss-mixture}) representing the slow (blue
    dash-dotted line) and fast (yellow dashed line) walking
    dynamics.}\label{fig:gauss-mixture}
\end{figure}
We now take into consideration the free-stream dynamic, i.e.\ 
we consider configurations in which only one pedestrian at a time 
moves on the staircase in either direction.
The average walking speed in upstream direction is 
$\avgn{v}_{stairs\uparrow} \approx 0.63\;$m/s, with 
$\avgn{v}_{stairs\downarrow} \approx 0.73\;$m/s 
for the downstream direction.
The observed $0.1\;$m/s speed difference between ascending and
descending well compares with the $10\%$ difference in
speed reported by~\cite{fruin-book-1971}, and with previous 
figures from literature (Table.~\ref{tab:free-stream-vel}). 
However, thanks to the vast collection of trajectories gathered 
during our tracking campaign, in this work we can take our analysis
beyond the average case.
In Fig.~\ref{fig:stats}c we report the probability distributions of
the free-stream velocity. The red curve represents the case 
$\trajds_{stairs \uparrow}$, while the green one the case 
$\mathcal{T}_{stairs\downarrow}$. For comparison, we also present 
a distribution of the walking speed on the train platform (brown curve), 
which was obtained from pedestrian trajectories measured by a commercial
pedestrian tracking setup installed in front of the staircase 
- similar to the setup used in~\cite{pouw-plos-2020, pouw-cd-2021}.
The walking speed on the staircase exhibits a skewed distribution,
with a pronounced right tail, representing cases of fast walking/running people.
As a result, average speed values are significantly higher than the modal ones,
with respectively $\mode{v}_{stairs\uparrow} \approx 0.48\;$m/s and 
$\mode{v}_{stairs\downarrow} \approx 0.58\;$m/s for ascending and descending.
On the platform, we observe the opposite behavior, with a left-tailed 
distribution, where due to pedestrian waiting for their train
the modal value ($\mode{v}_{\rm platform} \approx 1.23\;$m/s) 
is significantly higher than the average 
($\avgn{v}_{\rm platform} \approx 1.02\;$m/s). 
In order to quantify the ratio between slow and fast walkers, we take into consideration a Gaussian-mixture modeling for the probability distribution of the free-stream velocity for the different cases presented in Fig. 4c. This modeling approach is aligned with the findings of \cite{saberi-physa-2015} which demonstrated that a mixture of two Gaussian distributions provides a robust fit for pedestrian bidirectional velocity data. In order to allow for a simple physical interpretation, we consider a 2-component model:
\begin{equation}\label{eq:gauss-mixture}
  P(v) \sim \phi_s \mathcal{N} (\mu_s, \sigma_s) + \phi_f \mathcal{N} (\mu_f, \sigma_f) , \quad \phi_s + \phi_f  = 1,
\end{equation}
where $\mathcal{N}$ is a Gaussian distribution with mean $\mu_i$, variance $\sigma_i$ and weight $\phi_i$, and where the subscript $s$ (or $f$) is used to refer to the parameters for slow (or fast) walkers.
In Fig.~\ref{fig:gauss-mixture} we show the fit for the data shown in
Fig.~\ref{fig:stats}c using Eq.~\eqref{eq:gauss-mixture}. For the
descending case we fit a Gaussian-mixture model with parameters $P(\phi_s = 0.35,
\mu_s = 0.58\; \sigma_s = 0.1, \phi_f = 0.65
\mu_f = 0.8, \sigma_f = 0.27)$.
For the ascending case,
instead, the results of the fit yields $\phi_s = 0.48$, $\mu_s = 0.49\;$m/s, $\sigma_s = 0.08\;$m/s, $\phi_f = 0.52$, $\mu_f = 0.75\;$m/s, $\sigma_f = 0.25\;$m/s.

\paragraph{Eulerian Fields.}
The stairs-escalator system has relatively narrow side boundaries
(about 5 body diameters). In combination with the intermediate
landing, this yields dynamics that are not spatially uniform. We
consider this in terms of floor usage and of walking velocity
field. As pedestrians ascend or descend the stairs, and as they
traverse the landing, they adjust their speed accelerating and
decelerating. Here, we report Eulerian fields of space occupancy
(position probability distribution), average speed, and
acceleration, computed on the entire dataset $\trajds$.

In Fig.~\ref{fig:heatmaps}b, we report the probability distribution
of pedestrian positions, $\prob(\xv)$ (colormap: probability
iso-contour).  Consistently with Fig.~\ref{fig:stats}b, the escalators
area feature the highest occupancy probability, peaking on the
ascending escalator. A clear lane on the right-hand side of each
escalator is observable. This is generated by pedestrians standing on
escalators on the right while leaving the left side to people that
overtake.

In the bulk of the staircase area, trajectories align along three main
lanes. These lanes are especially pronounced along the
ascending/descending region, while they are dimmer in the landing
area. This can be explained considering the average walking speed
field (Fig.~\ref{fig:heatmaps}c), The average in $x$
direction of the speed field is reported as a function of $y$ in
Fig.~\ref{fig:heatmaps}e). We observe that the
average speed on the staircase is around $0.6\;$m/s, but in the
landing zone it increases up to $1\;$m/s. This velocity growth
renders the average permanence time on the landing smaller thus the
lower probability.  In Fig.~\ref{fig:heatmaps}d, we report
the field of acceleration in $y$ direction, $a_y$ (transversal average
in Fig.~\ref{fig:heatmaps}e). Consistently with the speed field, we
observe a strong acceleration about half meter before the landing and,
symmetrically, a deceleration that starts almost immediately after the
person sets foot on the landing. Finally, a net increase in velocity
is observable at the end of the ascending escalator, as people start
to walk when they are close to the surface.

\subsection{Density patterns and interpersonal distance: crowd compressibility}\label{sec:compressibility}

\begin{figure}[t]
  \centering  
  \begin{overpic}[width=.48\linewidth]{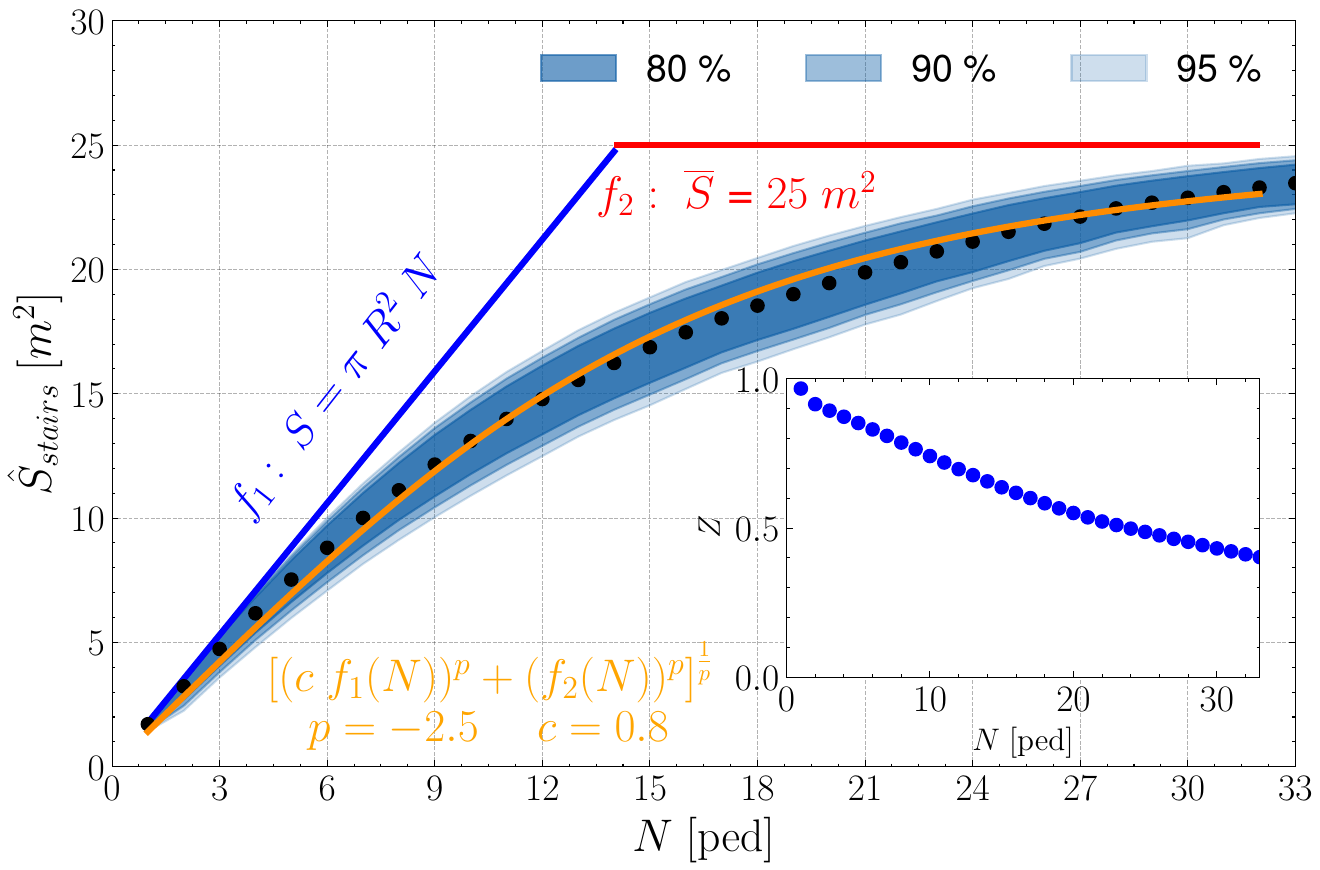}
    \put( 0.5, 60.0){(a)}
  \end{overpic}     
  \begin{overpic}[width=.48\linewidth]{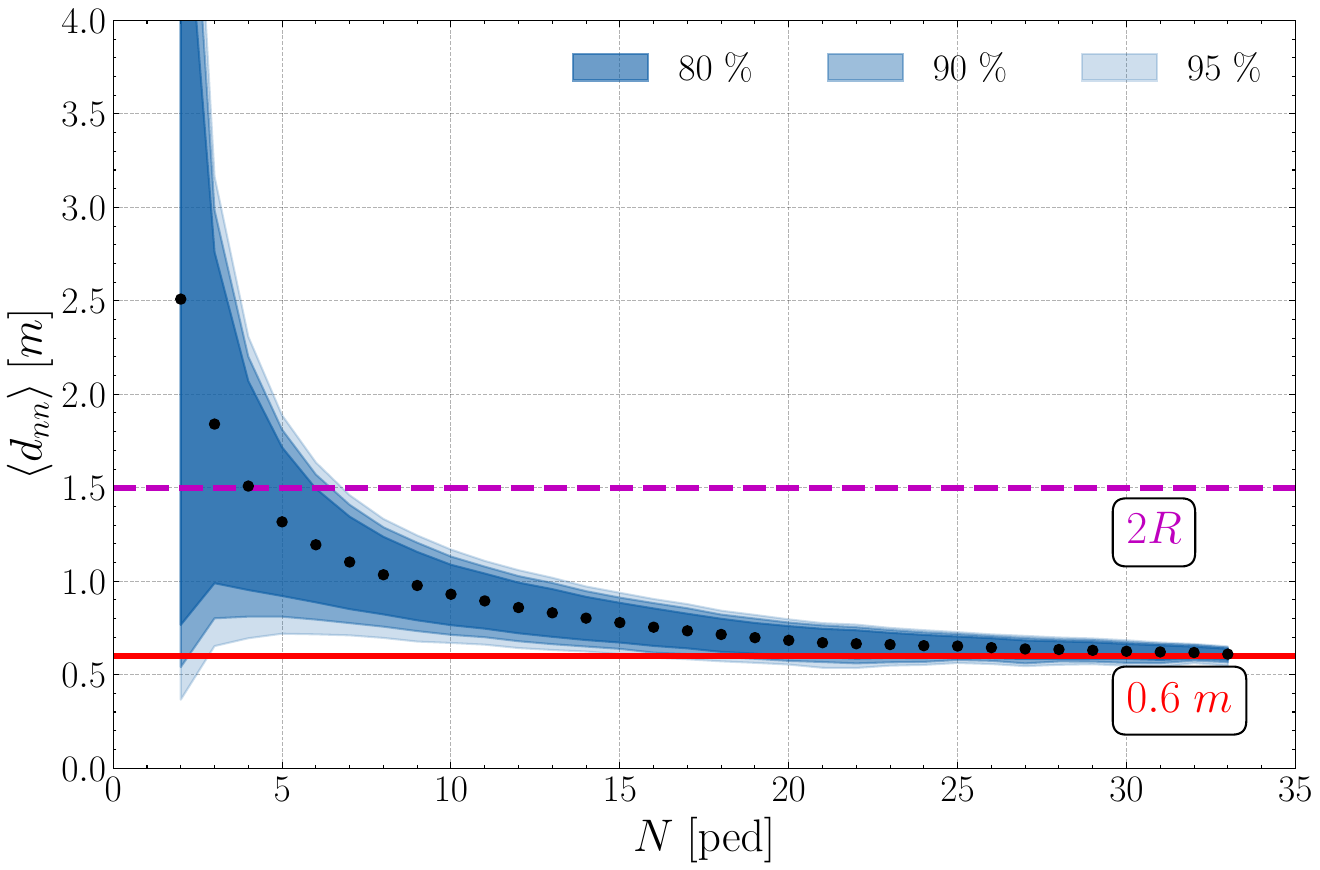}  
    \put( 0.5, 60.0){(b)}
  \end{overpic}           
  \caption{
    (a) Personal space union, $\hat S_{stairs}$, as a function of the
    number of pedestrians on the staircase. The average is represented
    with black dots and we show the $80^{th}$, $90^{th}$, and
    $95^{th}$ percentiles with different shades of blue. This provides
    insight in how the area is filled, with pedestrians accepting to occupy less than the available space, allowing for a compression. We superimpose the theoretical bounds to the occupied area, respectively $f_1(N) = \pi R^{2} N$ (blue line; all individual spaces do not intersect and are sufficiently far from the boundary) and the
    area of the staircase $f_2(N)=\bar{S} \approx 25 \;m^2$ (red line).
    While having initially a linear trend, $\hat S_{stairs}$ departs
    immediately from $f_1(N)$ and, for large $N$, the area approaches
    the global area. We report in yellow our model
    $\hat S^{model}_{stairs}(N)$ (Eq.~\eqref{eq:p-norm}), that through
    the single parameter $p=-2.5$ models the willingness of people of
    using less the space than available. In this respect, in the inset
    we report the compressibility factor $Z$ 
    (Eq.~\eqref{eq:Z-compressibility}) (b) Frame averaged mutual
    distance between closest neighbors $\langle d_{nn} \rangle$ as a
    function of the stairs occupation $N$. We observe a
    decrease in the mutual distance as the area gets filled with more
    pedestrians, converging to a minimal frame averaged mutual
    distance $d_{nn} \approx 0.6 \;m$.}\label{fig:density-def}
\end{figure}
In the previous section, we have highlighted the heterogeneous composition
of flow patters from data, showing how throughout the day one can observe
high density clusters alternating with low density to free-stream configurations. 
We discuss here how density effectively scales with the number of observed 
pedestrians and, as such, how the available area is filled. 

\begin{figure}[t]
  \centering
  \begin{overpic}[width=\linewidth]{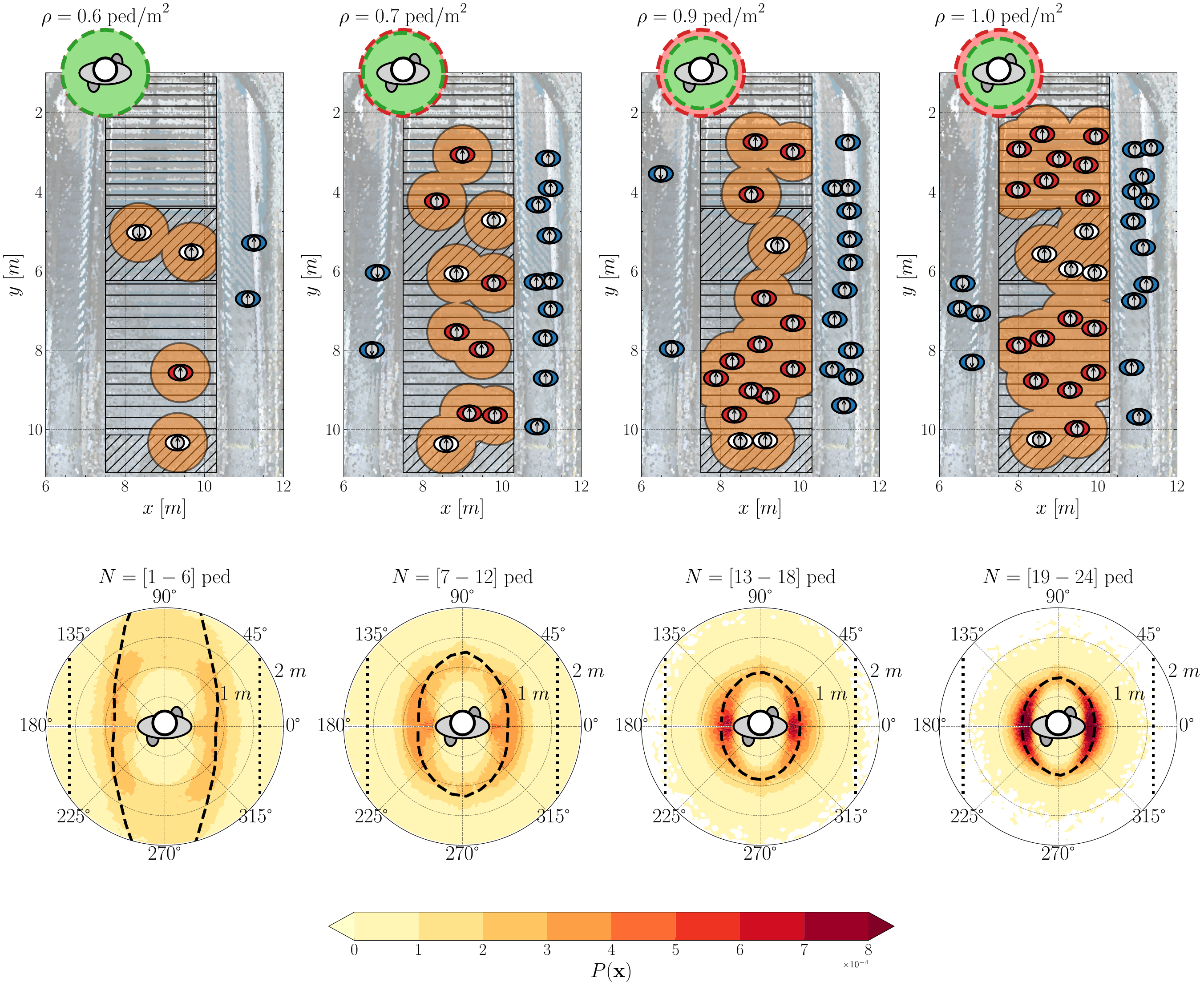}
    \put( 0, 80.0){(a)}
    \put( 24.0, 80.0){(b)}
    \put( 49.0, 80.0){(c)}
    \put( 74.0, 80.0){(d)}
    \put( 0, 35.0){(e)}
    \put( 24.0, 35.0){(f)}
    \put( 49.0, 35.0){(g)}
    \put( 74.0, 35.0){(h)}
  \end{overpic}
  \caption{ 
    (a-d) Overhead images of the staircase showing
    examples of the four increasing levels of stair occupation with
    respectively $4$, $10$, $15$, and $21$ pedestrians. Pedestrians
    located on the inclined plane of the staircase are colored in red,
    pedestrians located on a flat part of the staircase in white, and
    pedestrians on the escalators in blue. We indicate the occupied
    area from Eq.~\eqref{eq:area} with an orange color. On top of each
    panel we show the pedestrian density as calculated using
    Eq.~\eqref{eq:density-hat}. (e-h) Spatial probability distribution
    of nearest neighbor positions for the four increasing levels of
    stair occupation. We split our data set based on the stairs occupation in four evenly-sized bins with sides at $N=\{1,7,13,19,25\}$  pedestrians.
    A red color indicates high probability and a
    yellow color low probability. We report the average distance to
    nearest neighbors in a certain direction using a black dashed
    line. We observe an elliptical shape for low crowd density
    i.e.\ people allow others to be close in lateral direction but not
    in longitudinal direction. As the crowd density increases the
    shape becomes more circular and the radius decreases. To provide a
    sense of scale we impose two dotted black lines separated by
    $w=3.2\;$m, indicating the width of the staircase. Additionally,
    we impose an ellipse to indicate the typical body
    size. }\label{fig:nearest-neighbour-heatmap}
\end{figure}

In Fig.~\ref{fig:density-def}a we report the size of the occupied
area (Eq.~\eqref{eq:area}) as a function of the number of
pedestrians. This shows how the surface area on the staircase gets
filled due to how pedestrians choose to position themselves (see also examples in Fig.~\ref{fig:nearest-neighbour-heatmap}).
The area $\hat S_{stairs}$ is formed by the union of personal space
spheres from single individuals. Therefore, $\hat S_{stairs}$ admits
an upper bound given by the maximum between $N$ disjoint personal
space spheres (scaling as $\pi R^2 N$, blue line in
Fig.~\ref{fig:density-def}a), and the total (horizontal) surface of
the area of interest ($\bar{S} = \mbox{Area}(O_{region}) \approx 25~\rm{m^2}$, red line in
Fig.~\ref{fig:density-def}a). In formulas this reads
\begin{equation}\label{eq:area-max}
  \begin{cases}
    
    \hat S_{stairs}(N) \leq \hat S_{stairs}^{\max}(N) &= \min \{ f_1(N), f_2(N) \} \\
    &f_1(N) = \pi R^2 N \\
    &f_2(N) = \bar{S}
    \end{cases}
\end{equation}
In Fig.~\ref{fig:density-def}a, we report the average observed value of $\hat S_{stairs}(N)$, $\langle \hat S_{stairs}\rangle (N)$, and some percentiles. The average occupied area $\langle \hat S_{stairs}\rangle (N)$ grows linearly at small $N$ ($N < 6$), yet with a smaller slope than the upper bound: Eq.~\eqref{eq:area-max}
\begin{align}
  \label{eq:average-S-stairs-small-N}
 \langle \hat S_{stairs}\rangle (N) \approx c f_1(N)  && c\approx 0.8,\quad N < 6.
\end{align}
The constant $c<1$ is a consequence of multiple potential factors: on
one hand, we expect people to not strictly maintain a distance $d>2R$
even at small density - thus, yielding overlapping personal
spaces. This can be due, e.g., to small social groups, whose
individuals opt to walk in close proximity (see,
e.g., \cite{zanlungo-pre-2014}); note that in a similar Dutch train
station we measured about 15\%-20\% of the people to be in a group
\cite{pouw-plos-2020}).  Secondly, the definition itself of the
personal radius, $R$, has some degree of arbitrariness.  As $N$ grows,
the occupied area smoothly approaches the available surface. Thirdly,
our definition of occupied area is bounded by the observation region
$O_{region}$. This means that pedestrians close to the boundaries
experience a personal area smaller than the limit $\pi R^2$. Later in
the section, considering the dependencies on mutual distances, we
shall partially disentangle the roles of these factors.

The key observation here is that people willingly accept to occupy
less than full available area, accepting a ``compression''. Occupying
the full area available would mean having $\hat S_{stairs}(N)$ growing
linearly at small $N$ (accounting for possible arbitrariness in $R$),
up to the saturation at $\bar{S}$ that is reached with a kink point.
Note that this happens even though in diluted crowds pedestrians are
free to reposition themselves and, for instance, are able to overtake
slower walking pedestrians.

Conversely, the average (as well as high percentiles) of the occupied area depart rapidly from the theoretical bound and from a linear growth. We quantify this in terms of compressibility factor (see, e.g., for the definition of compressibility factor for gasses \cite{zucker-fundamentals-2019})
\begin{equation}
  \label{eq:Z-compressibility}
  Z = \frac{\hat S_{stairs}}{f_1(N)},
\end{equation}
i.e.\ the ratio between the occupied area and the maximum $f_1(N)$ assuming no overall geometric boundaries (inset in Fig.~\ref{fig:density-def}a). The observed compressibility factor rapidly decreases from $1$ as to indicate the presence of attractive interactions, which lower the area usage.

We further model this compressibility effect in terms of a power $p < 0$ of the $p$-norm between the two functions $cf_1(N)$ and $f_2(N)$:
\begin{equation}
  \label{eq:p-norm}
   \hat S_{stairs}^{model} (N) = \left[ (cf_1(N))^p + f_2(N)^p\right]^{\frac{1}{p}}.
 \end{equation}
Note that Eq.~\eqref{eq:p-norm}, in the limit $p\rightarrow -\infty$, converges to the theoretical bound (i.e.\ a linear trend connected with a kink to the saturated capacity, Eq.~(\ref{eq:area-max}) - this is a well known property of $p$-norms).  As $p$ grows from $-\infty$, the transition between $cf_1$ and $f_2$ becomes increasingly smoother  (e.g. $p=-1$ yields their harmonic mean). Considering $c\approx 0.8$ (cf. Eq.~\eqref{eq:average-S-stairs-small-N}), we found $p=-2.5$. We stress that this $p$ number encodes for a non-trivial crowd dynamics effect, specifically the willingness of pedestrians to accept a compression. We expect that the $p$-value will in principle depend on cultural differences, habits, geometry and crowd flow conditions.

\begin{figure}[t]
  \centering
  \begin{overpic}[width=\linewidth]{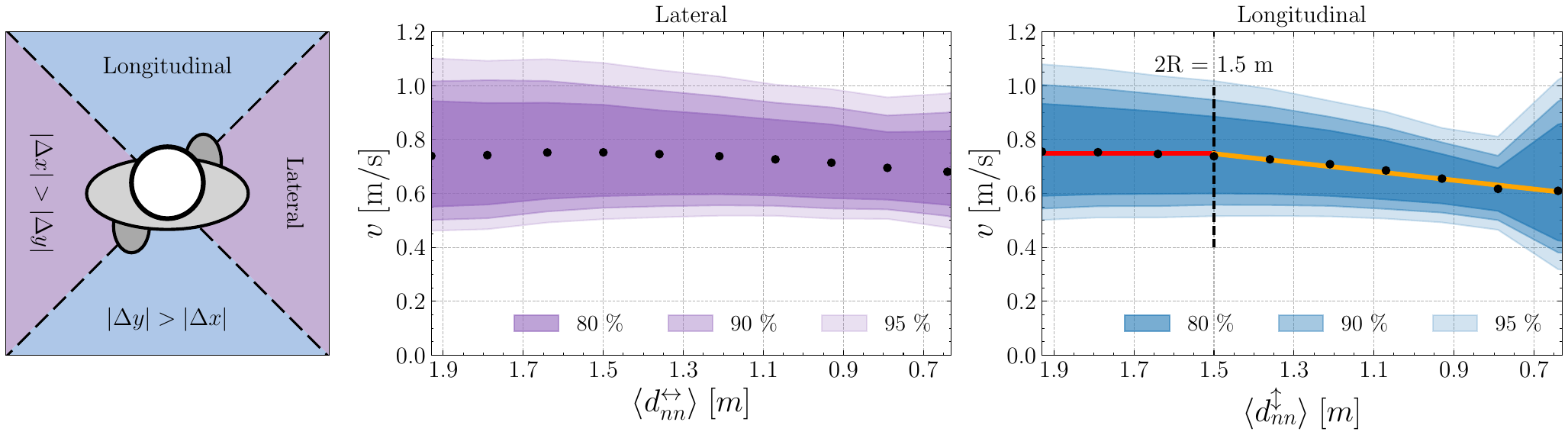}
    \put( -2.5, 25.0){(a)}
    \put(22.0, 25.0){(b)}
    \put(61.5, 25.0){(c)}
  \end{overpic}      
  \caption{(a) We consider a nearest neighbor to be positioned on the
    side if the transversal distance between the considered pedestrian
    and the neighbor is larger than the longitudinal direction
    ($|\Delta x_{ij}|>|\Delta y_{ij}|$, purple region). We consider a
    pedestrian in the front/back otherwise
    ($|\Delta x_{ij}|<|\Delta y_{ij}|$, blue region).  (b,c) Walking
    velocity, $v$, as a function of the frame averaged distance
    between closest neighbors restricting to a closest neighbor in transversal direction (b) and longitudinal direction (c).
    In (c) we fit for $\langle d_{nn}\rangle > 1.5\;$m a constant,
    $v(\langle d_{nn}\rangle)=0.75\;$m/s, with a red line and for
    $\langle d_{nn}\rangle < 1.5\;$m a linear fit otherwise (orange line, cf. Eq.~\eqref{eq:frontal-dependency}). We observe from (b) that a high lateral
    density does not impact the flow velocity and from (c) that the
    flow velocity decreases as the longitudinal distance between
    pedestrians gets less than $\langle d_{nn}\rangle =
    1.5\;$m. The velocity fluctuations decrease until $\langle d_{nn}\rangle \approx 0.8\;$m, for smaller closest neighbor distances (typically higher density) the fluctuation increase again. This likely connects with the necessity of avoidance/overtake maneuvers.}\label{fig:neighbor-distance}
\end{figure}

We conclude the section by considering the dynamics from the
perspective of single individuals analyzing how pedestrians arrange in space and how this influences the dynamics. Our key parameter
is the average distance to the nearest neighbor.  Specifically,
for every pedestrian, $i$, we measure on a frame-by-frame basis the
distance to their closest peer $j$:
\begin{equation}\label{eq:closest-d}
  d_{i,nn} = \min_{j}\left\{| \mathbf{x}_{i}-\mathbf{x}_{j}|\right\}.
\end{equation}
We then retain the frame-by-frame averages values of $d_{i,nn}$:
\begin{equation}\label{eq:neighbor}
  \avgn{d_{nn}} = \frac{1}{N_{\textrm{frame}}}\sum_{i=1}^{N_{\textrm{frame}}}{d_{i,nn}}.
\end{equation}
In Fig.~\ref{fig:density-def}b, we report the frame-averaged
distance $\avgn{d_{nn}}$ as a function of stairs occupation $N$.
For $N \leq 3$ the average nearest neighbor distance remains well above $2R$ with, however, large fluctuations. This supports the presence of social groups ($\avgn{d_{nn}}\approx R$), but also the fact that pedestrians opt to walk close to boundaries even at low densities (in fact this yields a reduction $\hat S_{stairs}$ from the upper bound $f_1(N)$ due to the fact that only personal areas within $O_{region}$ are considered). On the other hand, as the density increases, the average distance approaches a limiting
value $\avgn{d_{nn}} \approx 0.6\;$m. Note that two threads of the staircase measure $0.6\;$m.

We conclude the section by considering how nearest neighbors arrange in space and how this affects the dynamics.
For each pedestrian $i$, we measure frame-by-frame 
the relative position of the nearest neighbor.
In Fig.~\ref{fig:nearest-neighbour-heatmap}, we present heatmaps of
the spatial distribution for the location of nearest neighbors for 
increasingly large values of the local density.
In diluted conditions we observe low probabilities for finding the
nearest neighbor aligned with the direction of traveling.
On the other hand the probability is maximum along the lateral
direction, for a distance $d \approx 0.9\;$m. 
In the Figure we report with a black dotted line the iso-contour 
given by the most probable value $\avgn{d_{nn}} \approx 0.6\;$m. 

In Fig.~\ref{fig:nearest-neighbour-heatmap}(bottom panels), we present
heatmaps of the spatial distribution for the location of nearest
neighbors for increasingly large values of the local density (examples
of pedestrians distribution at the related density values are in the
top panes). We report with a black dotted line the 1d-manifold given
by the most probable
location of nearest neighbors in every direction. This yields an
elliptical shape, with major axis aligned along the pedestrian walking
direction, but with decreasing eccentricity as the local density is
increased.
In diluted conditions, we observe that closest neighbors have maximum probability to be located along the sides ($d_{side} \approx 0.9\;$m), conversely plenty of headway ($d_{headway} > 2.5\;$m) is left in frontal direction. As density increases the manifold of the average nearest neighbor position contracts approaching the $d \approx 0.6\;$m, consistently with the asymptotic trend of Fig.~\ref{fig:density-def}.

As one can expect (e.g., \cite{rudina-applsci-2022}), the
position of the closest neighbor influences pedestrians in an
anisotropic way. To conclude the section, in
Fig.~\ref{fig:neighbor-distance}, we report the frame-averaged walking
velocity (Eq.~\eqref{eq:walking-velocity}) as a function of the average
distance between nearest neighbors (Eq.~\eqref{eq:neighbor}). Note that
here we make no distinction between pedestrians ascending and
descending the stairs. We separate the condition in which the nearest
neighbor is positioned on the side
($|\Delta x_{ij}| > |\Delta y_{ij}|$, cf. Fig.~\ref{fig:neighbor-distance}a), and in the front/back
($|\Delta x_{ij}| < |\Delta y_{ij}|$). To avoid confusion, 
we indicate these distances as
$\avgn{d_{nn}^{\leftrightarrow}}$ and $\avgn{d_{nn}^{\updownarrow}}$,
and report the velocity in dependence on these quantities
respectively in Fig.~\ref{fig:neighbor-distance}b
and~\ref{fig:neighbor-distance}c.
From Fig.~\ref{fig:neighbor-distance}b, we observe that a closest neighbor 
on the lateral side has practically no influence on the average velocity, i.e.\ 
\begin{align}
  \label{eq:lateral-dependency}
  v\left(\avgn{d_{nn}^{\leftrightarrow}}\right)\approx 0.75\;\mbox{m/s} \qquad\forall \avgn{d_{nn}^{\leftrightarrow}}.
\end{align}
Conversely, $v$ shows strong variations when the closest neighbor approaches 
in longitudinal direction reducing the headway (Fig.~\ref{fig:neighbor-distance}c). 
In particular, the  velocity $0.75\;\mbox{m/s}$ (cf. Eq.~\eqref{eq:lateral-dependency}) 
holds only for $\avgn{d_{nn}^\updownarrow} < 1.5\;\mbox{m}$, and decreases otherwise. 
Overall, we observe the following piece-wise approximated dependency
\begin{equation}
  \label{eq:frontal-dependency}
  v\left(\avgn{d_{nn}^{\updownarrow}}\right)\approx
  \begin{cases}
    0.75\;\mbox{m/s} & \avgn{d_{nn}^{\updownarrow}} > 1.5 \;\mbox{m} \approx 2R \\
    0.16 \cdot \avgn{d_{nn}^{\updownarrow}} + 0.5 &  \avgn{d_{nn}^{\updownarrow}} < 1.5 \;\mbox{m} \\
    \mathrm{Units}\quad v :\;[\mathrm{m/s}] \quad \avgn{d_{nn}^{\updownarrow}} : \;[\mathrm{m}].
  \end{cases}  
\end{equation}
Notice that the transition around  $1.5 \;\mbox{m} \approx 2 R$   supports our scale for the definition of the size of the personal space (Sec.~\ref{sec:density-vel}).

In Fig.~\ref{fig:neighbor-distance}b, we further notice that the velocity fluctuation decreases until $\avgn{d_{nn}^{\updownarrow}} \approx 0.8\;$m, to then increase  substantially. This likely connects with the fact that as the pedestrian in front gets too close, evasive maneuvers become necessary.

\subsection{Probabilistic fundamental diagrams}\label{sec:probab-fund-diag}

\begin{figure}[t]
  \centering 
 \begin{overpic}[width=\linewidth]{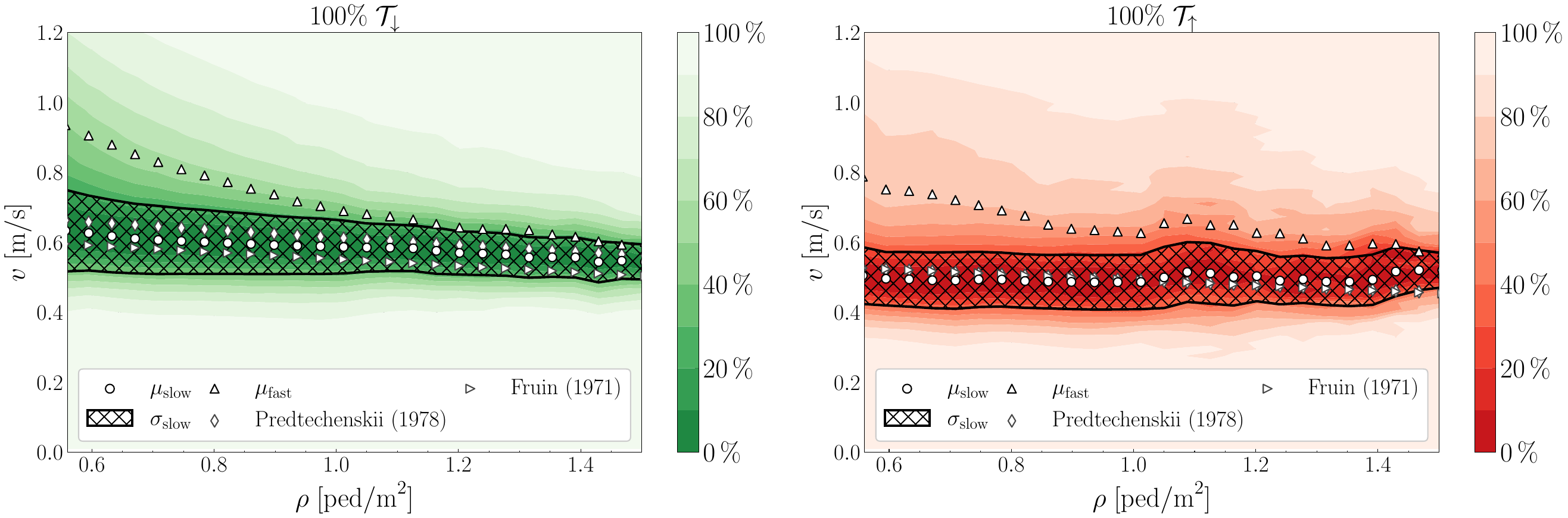}
    \put(-1.5, 30.0){(a)}
    \put(50.0, 30.0){(b)}
  \end{overpic}
  \caption{Fundamental diagrams for (a) unidirectional flow
    descending, $100\% \;\mathcal{T}_{\downarrow}$, and (b)
    unidirectional flow ascending, $100\%
    \;\mathcal{T}_{\uparrow}$ reporting the walking velocity as a
    function of the crowd density. We use a colormap to show the
    confidence intervals in our data set. We include the average
    walking speed of slow walkers $\mu_{slow}$ (circles) with a
    hatched domain to show one standard deviation to the mean of slow
    walkers $\sigma_{slow}$. Additionally, we report the average
    walking speed of fast walkers $\mu_{fast}$ (upward pointing
    triangles), and for comparison the FD reported by
    \cite{fruin-book-1971} (right pointing triangles) and by \cite{predtechenskii-report-1978} (diamonds).
  }\label{fig:fd-ps}
\end{figure}

\begin{figure}[t]
  \centering  
  \begin{overpic}[width=.99\linewidth]{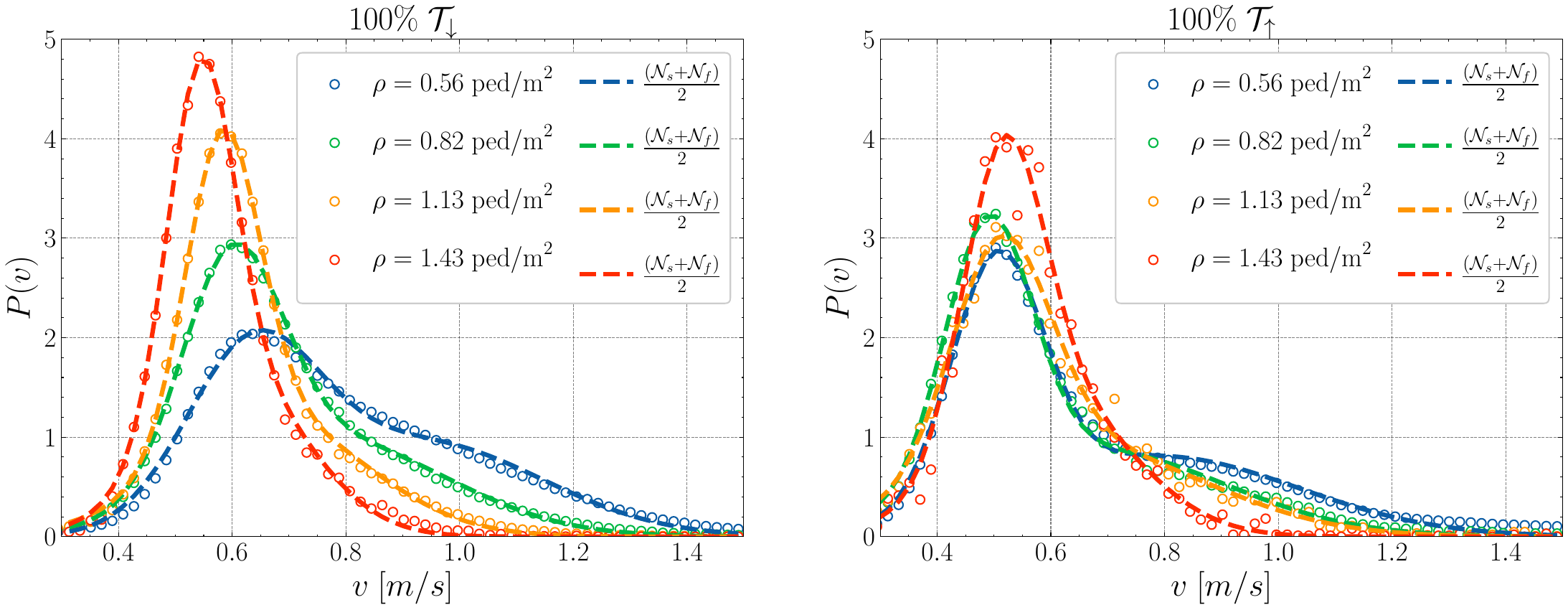} 
    \put(-1.5, 35.0){(a)}
    \put(50.0, 35.0){(b)}
  \end{overpic}
  \begin{overpic}[width=.99\linewidth]{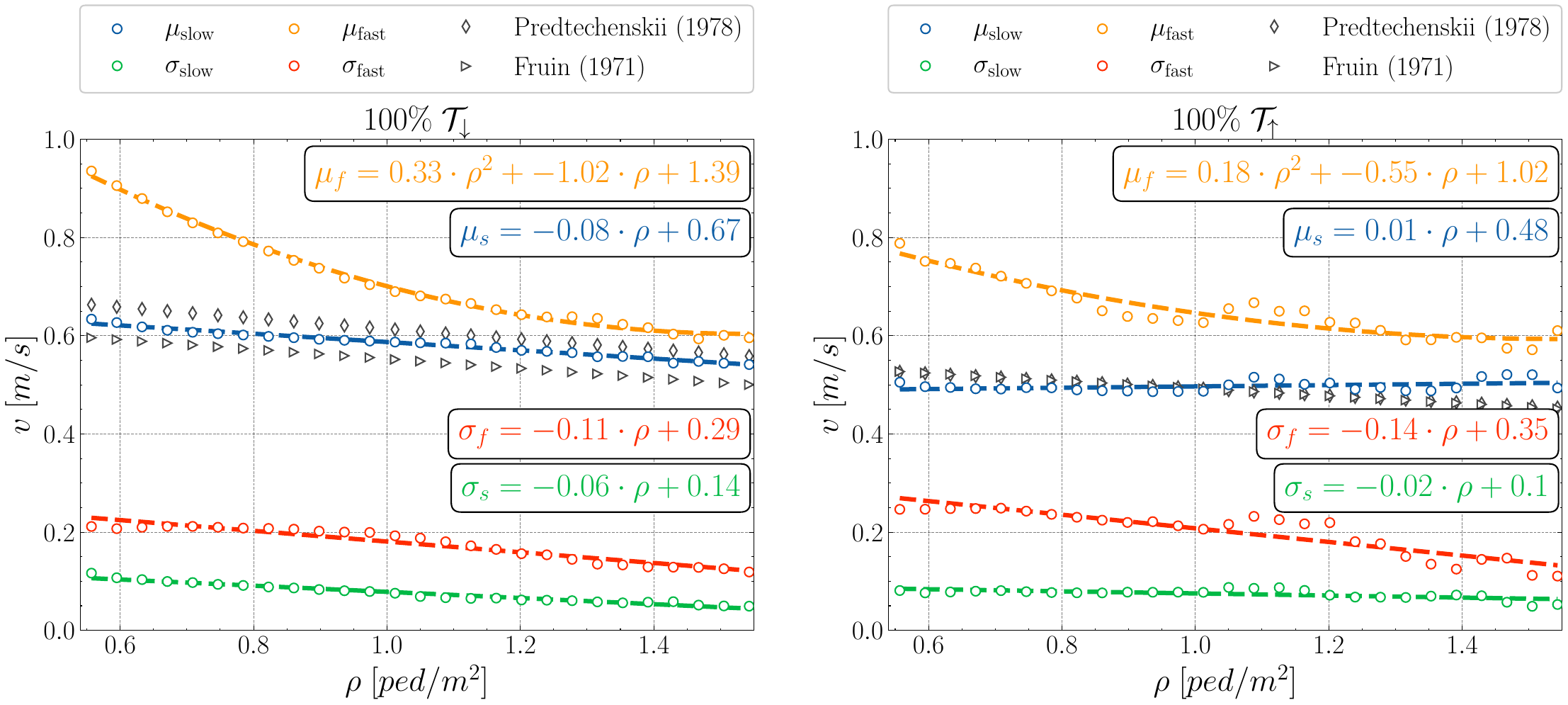} 
    \put(-1.5, 35.0){(c)}
    \put(50.0, 35.0){(d)}
  \end{overpic}
  
  \caption{(a,b) Probability distribution function of the walking
    velocity on the staircase at crowd densities
    $\rho = [0.56, 0.82, 1.13, 1.43]\; \mathrm{ped/m^2}$ for (a)
    descending and (b) ascending pedestrian movement. Every
    distribution is fitted with a Gaussian mixture with equal weights
    (Eq.~\eqref{eq:gauss-mixture-equal}). (c,d) the Gaussian mixture
    fit parameters $\mu$ and $\sigma$ across the observed density
    range for the slow and fast (c) downstairs and (d) upstairs
    walking movements. We employ a linear fit to $\mu_{slow}$,
    $\sigma_{slow}$ and $\sigma_{fast}$ and a quadratic fit to
    $\mu_{fast}$. We include the fundamental diagram by
    \cite{fruin-book-1971} and \cite{predtechenskii-report-1978}
    (Eq.\eqref{eq:fruin-model-des} and Eq.\eqref{eq:fruin-model-asc})
    for comparison.}\label{fig:fd-param-fitting}
\end{figure}

In this section, we employ our trajectory dataset to report and
compare fundamental diagrams (FD) (density-velocity relations) for
pedestrians ascending and descending our staircase.  We specifically
consider probabilistic counterparts of fundamental diagrams, that
include average (and modal) velocities, common fluctuations (fractions
of one standard deviation), and rare fluctuations (percentiles of the
speed distributions conditioned to the density).
It is clear that only very high statistics datasets can serve to this purpose. 
In Fig.~\ref{fig:fd-ps} we present the fundamental diagram for
the walking velocity, as function of the pedestrian density on the
staircase. We restrict our analysis to unidirectional flows, showing
in panel (a) the case where all pedestrians descend the staircase, with
the ascending case shown in panel (b).
In each plot, we show with a colormap the confidence intervals of our
data for given densities i.e.\ the percentage of recordings at that density that falls within the interval. For comparison, we also include in the plot the FD
by \cite{fruin-book-1971} and \cite{predtechenskii-report-1978}, which
can be parametrized with the following expressions (from
\cite{chen-ft-2018})
\begin{equation}\label{eq:fruin-model-des}
  v(\rho)_{\rm fruin} = -0.097\cdot\rho + 0.65, \quad v(\rho)_{\rm predtechenskii} = 0.009\cdot\rho^2- 0.125\cdot\rho + 0.73 \quad
\end{equation}
\begin{equation*}
  \mathrm{Units}\quad \rho:\;[\mathrm{ped/m}^2] \quad v:\;[\mathrm{m/s}]
\end{equation*}
for the descending case, and 
\begin{equation}\label{eq:fruin-model-asc}
  v(\rho)_{\rm fruin} = - 0.077\cdot\rho + 0.57, \quad v(\rho)_{\rm predtechenskii} = 0.003\cdot\rho^2- 0.08\cdot\rho + 0.57 \quad
\end{equation}
\begin{equation*}
  \mathrm{Units}\quad \rho:\;[\mathrm{ped/m}^2] \quad v:\;[\mathrm{m/s}]
\end{equation*}
for the ascending case.

Finally, we report in the diagram the average values of the
parametrization of the velocity distribution function (for a fixed
density value), which was established in Eq.~\eqref{eq:gauss-mixture},
and which allows to treating the average walking speed of slow and
fast walkers separately.

From Fig.~\ref{fig:fd-ps}a, we observe that average walking speed of
slow walkers $\mu_{slow}$ falls in between the two curves in
Eq.~\eqref{eq:fruin-model-des}, taking an average value of
$\approx 0.6\;$m/s, almost independent with respect to the local
density. On the other hand, for fast walkers ($\mu_{fast}$), we
observe a significant decreasing trend, starting from a free-flow
value of $\approx 1.0\;$m/s which converges to $\mu_{slow}$ for larger
values of the local density (i.e.\ in this case we cannot really
distinguish fast and slow walkers).
This pattern emerges also looking at fluctuations around the mean,
which decrease at large density values, as shown by the colormap.

Similar considerations apply to Fig.~\ref{fig:fd-ps}b, where we
consider, instead, the case of pedestrians walking in ascending
direction.
Here, the difference between slow and fast walkers is more
pronounced at large density values, although fewer data is available in
this case (see Fig.~\ref{fig:stats}d).

We conclude by providing a compact parametrization of the data
presented in Fig.~\ref{fig:fd-ps}, to aid future comparisons,
and especially, to allow using the information reported here in
realistic simulations of pedestrian flows on staircases.
We consider once again Eq.~\eqref{eq:gauss-mixture}, this time for
simplicity we use equal weights for the Gaussian distribution
associated to slow and fast walkers, i.e.:
\begin{equation}\label{eq:gauss-mixture-equal}
  P(v) \sim \frac{1}{2} \left( \mathcal{N} (\mu_s, \sigma_s) + \mathcal{N} (\mu_f, \sigma_f) \right) \quad .
\end{equation}
Next, we use Eq.~\eqref{eq:gauss-mixture-equal} to fit the (empirical)
distribution function of pedestrian walking velocity for fixed values
of density.
We report a few examples in Fig.~\ref{fig:fd-param-fitting}a and
Fig.~\ref{fig:fd-param-fitting}b, for flows in descending and ascending
direction respectively, which show that the approximation
$\phi_{s} = \phi_{f} = 1/2$ allows for an excellent fit of the PDFs.
We conclude taking a further step, which consists in fitting the parameters of 
the two Gaussian distributions as function of the local density $\rho$.
The results of the fit, reported in Fig.~\ref{fig:fd-param-fitting}c-d, delivers
the following expressions:
\begin{equation}\label{eq:fit-ascent}
  \begin{array}{lrlr}
    \mu_f =& 0.33 \rho^2 - 1.02 \rho + 1.39, \quad & \sigma_f =& - 0.11 \rho + 0.29 \\
    \mu_s =&             - 0.08 \rho + 0.67, \quad & \sigma_s =& - 0.06 \rho + 0.14 \\
  \end{array}
\end{equation}
\begin{equation*}
      \rm{Units}\quad \rho:\;[\rm{ped/m}^2] \quad \mu_{s/f}:\;[\rm{m/s}] \quad \sigma_{s/f}:\;[\rm{m/s}]
\end{equation*}
for the descending case, and
\begin{equation}\label{eq:fit-descent}
  \begin{array}{lrlr}
    \mu_f =& 0.18 \rho^2 - 0.55 \rho + 1.02, \quad & \sigma_f =& - 0.14 \rho + 0.35 \\
    \mu_s =&               0.01 \rho + 0.48, \quad & \sigma_s =& - 0.02 \rho + 0.10
  \end{array}
\end{equation}
\begin{equation*}
      \rm{Units}\quad \rho:\;[\rm{ped/m}^2] \quad \mu_{s/f}:\;[\rm{m/s}] \quad \sigma_{s/f}:\;[\rm{m/s}]
\end{equation*}
for the ascending case.

\section{Discussion}\label{sec:conclusions}
In this work, we investigated with unprecedented statistical
resolution the dynamics of pedestrians as they ascend or descend a
large staircase in a railroad station (Eindhoven Central Station, The
Netherlands). By employing a state-of-the-art pedestrian tracking
system based on a grid of overhead depth sensors and hinging on the
latest computer vision algorithms, we have recorded over 3 million
individual trajectories under various flow conditions (unidirectional
and bi-directional with various level of mixing), in an unbiased and
privacy-respectful manner. The dataset that we collected is at least
three orders of magnitude larger in terms of data volume than datasets
currently considered in the literature. This is key in order to study
pedestrian dynamics beyond the averages in terms of fluctuations. Our
experimental dataset is characterized by extremely high
accuracy, with an F1 Score on localization always above $96\%$ and
above $99\%$ for less than $0.95$ ped/m$^2$, thanks to a dedicated
hand-annotated training set and effective data
augmentation. Due to its real-life nature the dataset is limited to crowd densities not exceeding 1.5 ped/m$^2$.

We provided a phenomenological analysis of pedestrian dynamics
considering various key components.  First, we investigated spatial
fields of positions, velocity and accelerations.  For a staircase as
ours, pedestrians would arrange with the highest probability along
three parallel lanes. These lanes, partially diffuse on the
intermediate landing due to speed changes.  The velocity and
acceleration fields show that the walking velocity is non-uniform on
the staircase.  In free flow, we observed velocities, on average, of
$0.63\;$m/s in ascending direction and $0.73\;$m/s while descending.
These readings have an increment larger than $50\%$ (reaching
$1.1\;$m/s) when people step on the intermediate landing.  The walking
velocity on the landing remains however significantly lower than what
observed on the rest of the platform (mode: $1.23\;$m/s).  This is
likely due to the short length of the landing ($1.5\;$m) that does not
allow for reaching a comfortable speed.

Secondly, we have investigated how pedestrians fill the available
space. We considered a definition of density hinged on assigning to
each pedestrian a personal space (circular, with radius $R=0.75\;$m).
We considered the total occupied area as the union of these personal
spaces. We defined as density the ratio between the number of
observed people and such area. Our choice of $R$ yields a
level-of-service for well separated pedestrians at the interface
between level A (free-flow) and B (slightly restricted flow). This choice of $R$ is however consistent with the sudden velocity reduction that we observe as a closest neighbour in front gets to a distance smaller than $2R$.

The total occupied area admits an upper bound defined by the (minimum between) the area of $N$ disjoint personal areas (linear growth) and the total surface of the observed region. Notably pedestrians opt not to fill all the available space (differently, e.g., from a ideal gas) and even at relatively low density levels they occupy a smaller area than the concept of personal space would predict. It is interesting to notice that the occupied area immediately departs from a linear growth and approaches the total area of the facility only smoothly, as the number of pedestrians $N$ grows. This is due to an interplay of different elements, including the presence of social groups (yielding people walking in proximity). We interpret this in terms of a compressibility factor, $Z$, of the crowd being smaller than $1$ to represent attractive interactions (in analogy with similar effects in non-ideal gases). Most importantly we introduce a novel parameter, $p$-value, that allows to  quantitatively model the way pedestrians fill the available space. In the specific setting studied here we find that the area gets filled consistently with a value $p=-2.5$. This quantity expresses how smoothly the saturation capacity is reached (note that $p=-\infty$ would yield an ideal gas-like behavior for which all the available space is used and pedestrians do not accept to compress until needed). In general we expect the $p$-value to correlate at least with cultural preferences, with the specific geometry, with the crowd composition and with the specific flow conditions. We believe that a deeper understanding of the $p$-value and its usage in crowd simulations will be key to reproduce realistic crowd dynamics scenarios and risk assessments.

Considering how the distance to the closest neighbors changes with
density we observed that pedestrians strive to maintain a distance to
their closest neighbor that is quasi elliptical in shape.
We observed besides that the presence of pedestrians on closed proximity on the side would not influence the walking speed, that instead starts to diminish as the person in front gets closer than approximately 2$R$. As a consequence, during our year-long measurement campaign we rarely observe configurations with density values larger than 1.5 ped/m$^2$, at variance with data from lab experiments and evacuation drills. We stress that the data collected during this period accurately reflects the typical traffic at a train station in the busiest railway network of Europe.
  
Finally, we have investigated the relationship between density and
velocity under diverse flow conditions exploring complete probability distributions. We have shown that a key component for an accurate modeling of the probability, due to the presence of a significant tail at high velocity, is the consideration of a mixture of two Gaussian distributions. In both cases of ascending and descending pedestrians we have provided a linear parametrization of how the mean and the variance of these two components scale with density. We deem these parametrizations as key reference towards  more  accurate crowd simulation models and facility design.

Probabilities are vital in performance-based engineering, guiding the design of infrastructures to withstand potential extreme events. Our study facilitates the evaluation of staircase performance indicators, such as pedestrian flux, across a spectrum of probabilities. For instance, at probability 1\%, 5\%, and 10\%, we expect flux reductions of 50\%, 35\%, and 30\%, respectively. Capacity reductions of this kind can raise safety concerns with potential financial implications. This type of insights, attainable only through extensive real-life studies, underline the significance of high-statistics real-world measurements in engineering.

In future works, we plan to investigate the formation of queues 
at the top of escalators and staircase, which originate as people 
try to establish a comfort zone in order to safely descent in 
crowded configurations, and to establish stochastic quantitative models for the individual and ensemble dynamics. 

\section*{Acknowledgments}
This work is supported by the HTSM research program ``HTCrowd: a
high-tech platform for human crowd flows monitoring, modeling and
nudging'' with project number 17962, financed by the Dutch Research
Council (NWO). We acknowledge Gijs Mescher for his valuable
contribution in the sensor calibration and fusion of depth images.

\appendix
\section*{Appendix}

\section{YOLO for pedestrian localization on depth images}\label{app:yolo}

To the best of our knowledge, there are currently no publicly available datasets containing overhead pedestrian depth images suitable for our specific use case. Consequently, we constructed our own dataset to train the YOLO model. In this section we describe the process used to create and validate this dataset.

\subsection{Dataset creation}

\begin{figure}[h]
  \begin{overpic}[width=.49\linewidth]{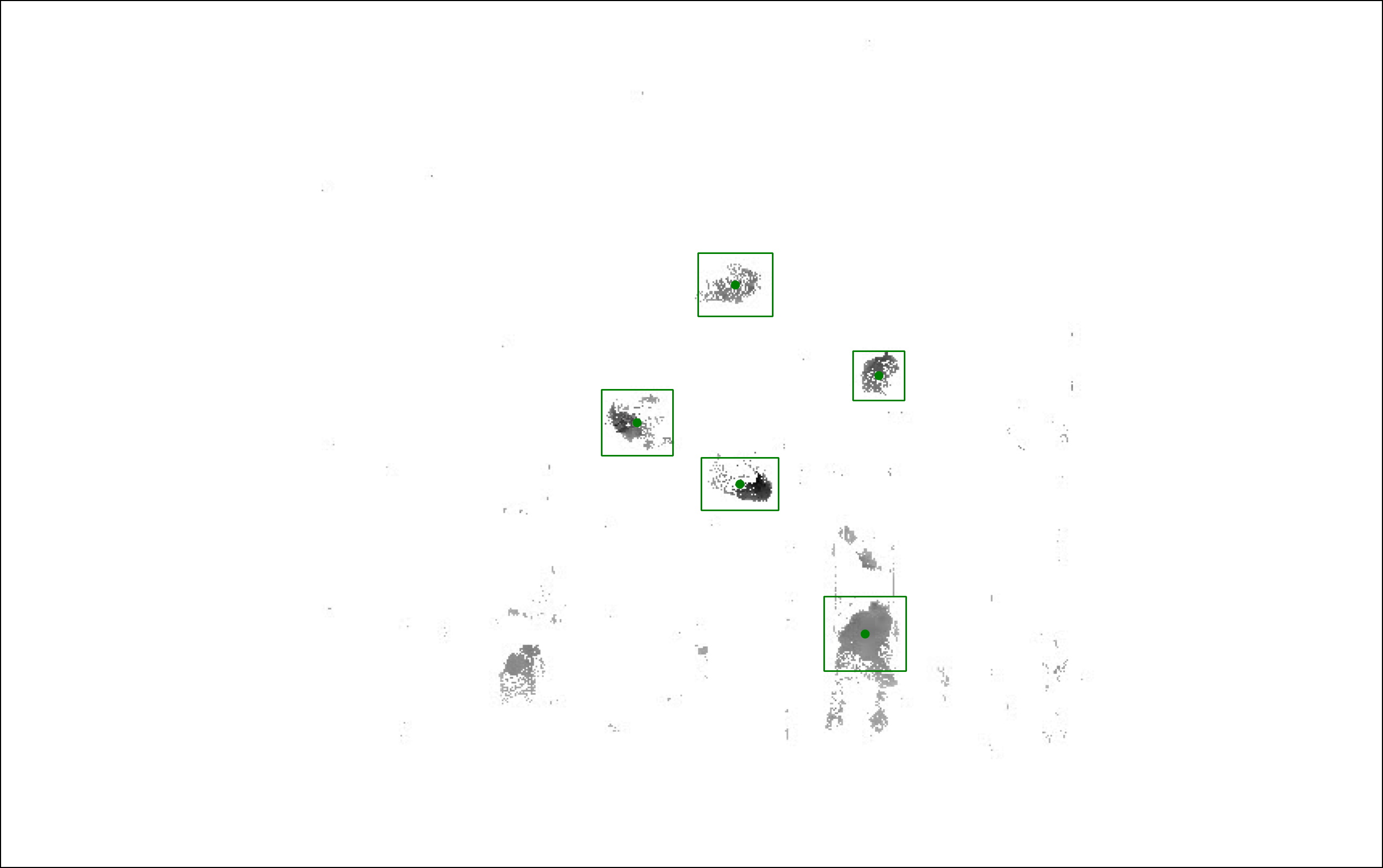} 
    \put(1, 2.0){(a)}
  \end{overpic}
  \begin{overpic}[width=.49\linewidth]{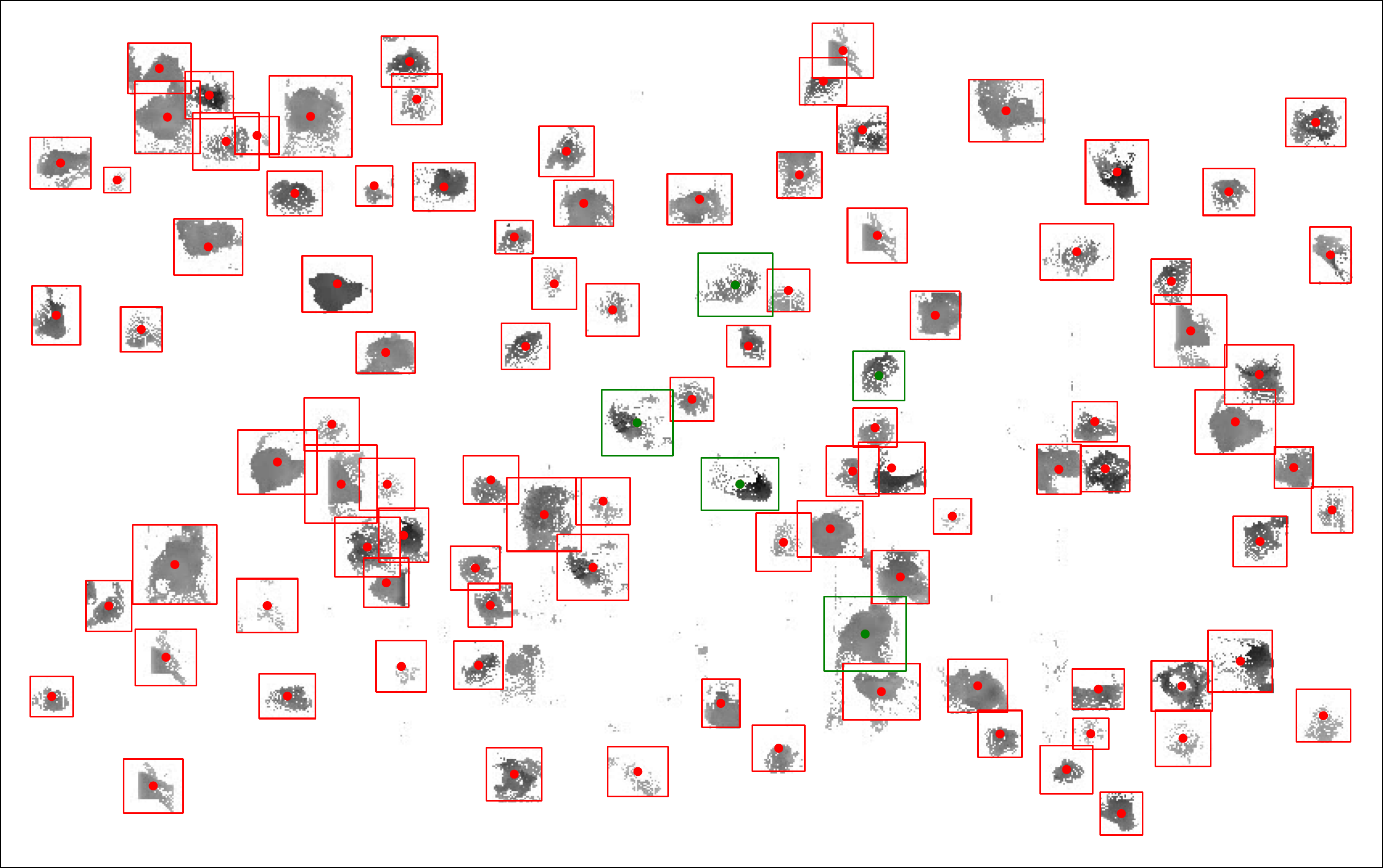} 
    \put(1, 2.0){(b)}
  \end{overpic}

  \caption{Example of an image in our CutMix-like approach to create an image for our training dataset. The green bounding boxes represent the original annotations, while the red bounding boxes represent the inserted reference annotations. (a) Original image with annotations. The non-annotated gray spots is noise that is present in the data. (b) Image with inserted reference annotations.}\label{fig:dataset-generation}
\end{figure}

We generate the dataset using an active learning cycle
\cite{settles-2009-active-learning}. Initially, we manually annotated
approximately 100 pedestrians in 15 frames with low global
densities. We crop the annotations at the edge of the bounding box and
store these annotations as references. Subsequently, we insert a
randomized quantity of these references into frames with existing
annotations using a CutMix-like approach, allowing for a slight
overlap (cf. \cite{yun-arxiv-2019-cutmix}). By allowing this slight
overlap, we aim to generate new and distinctive shapes that are
similar to people being in proximity to each other, effectively
augmenting our dataset. An example of this process can be seen in
Fig.~\ref{fig:dataset-generation}. On this basis, we generate about
$16000$ images covering a range of density levels. These images serve
as the training data set for the first training iteration of our
YOLOv7 model.

\begin{figure}[p]
  \centering
  \begin{overpic}[width=.3\linewidth]{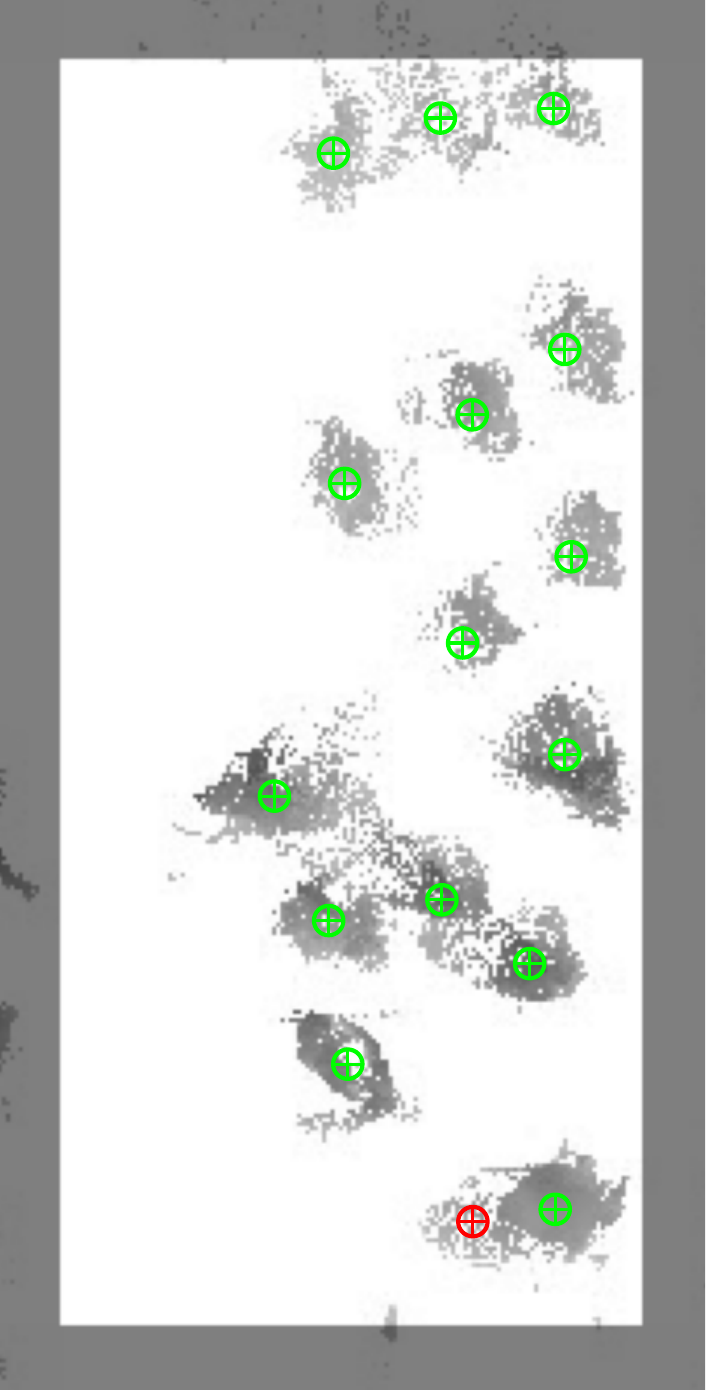} 
    \put(5, 92){(a)}
    \put(5, 97){\color{white}$0.60$ ped/m$^2$}
  \end{overpic}
  \begin{overpic}[width=.3\linewidth]{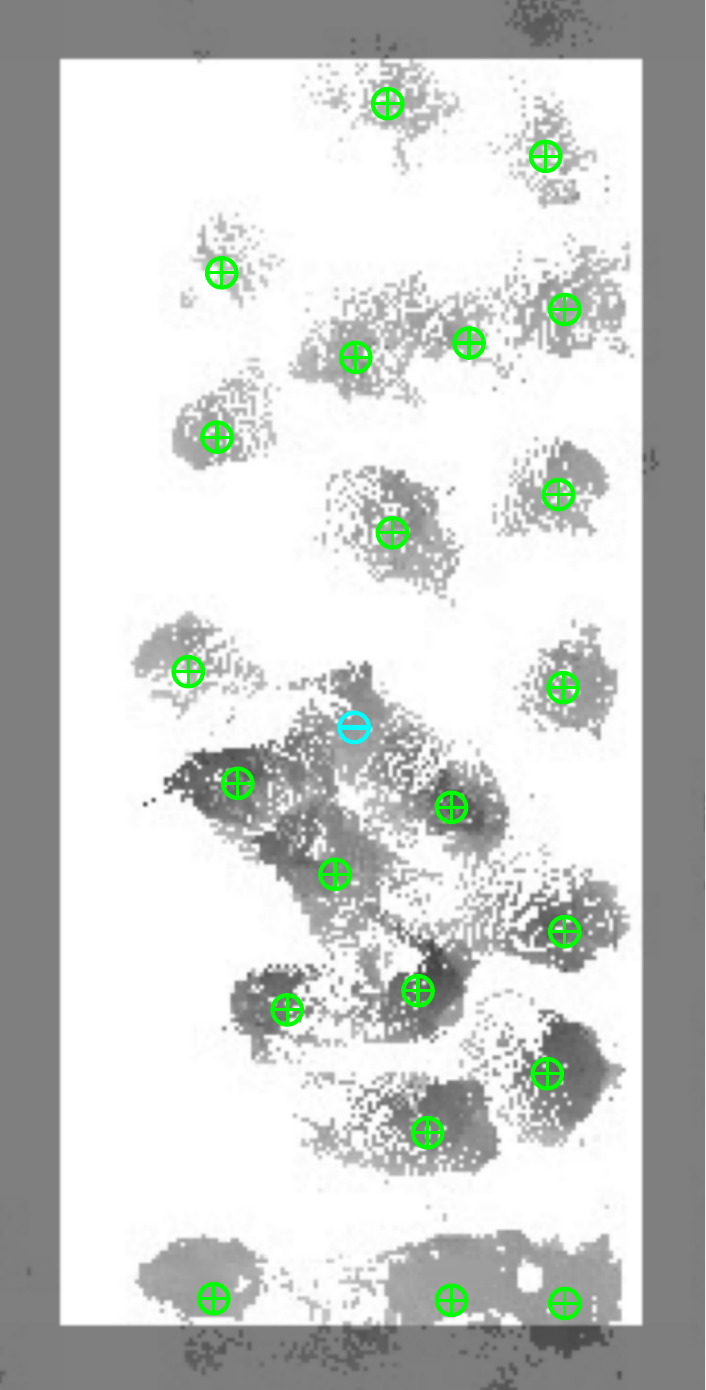} 
    \put(5, 92){(b)}
    \put(5, 97){\color{white}$0.92$ ped/m$^2$}
  \end{overpic}
  \begin{overpic}[width=.3\linewidth]{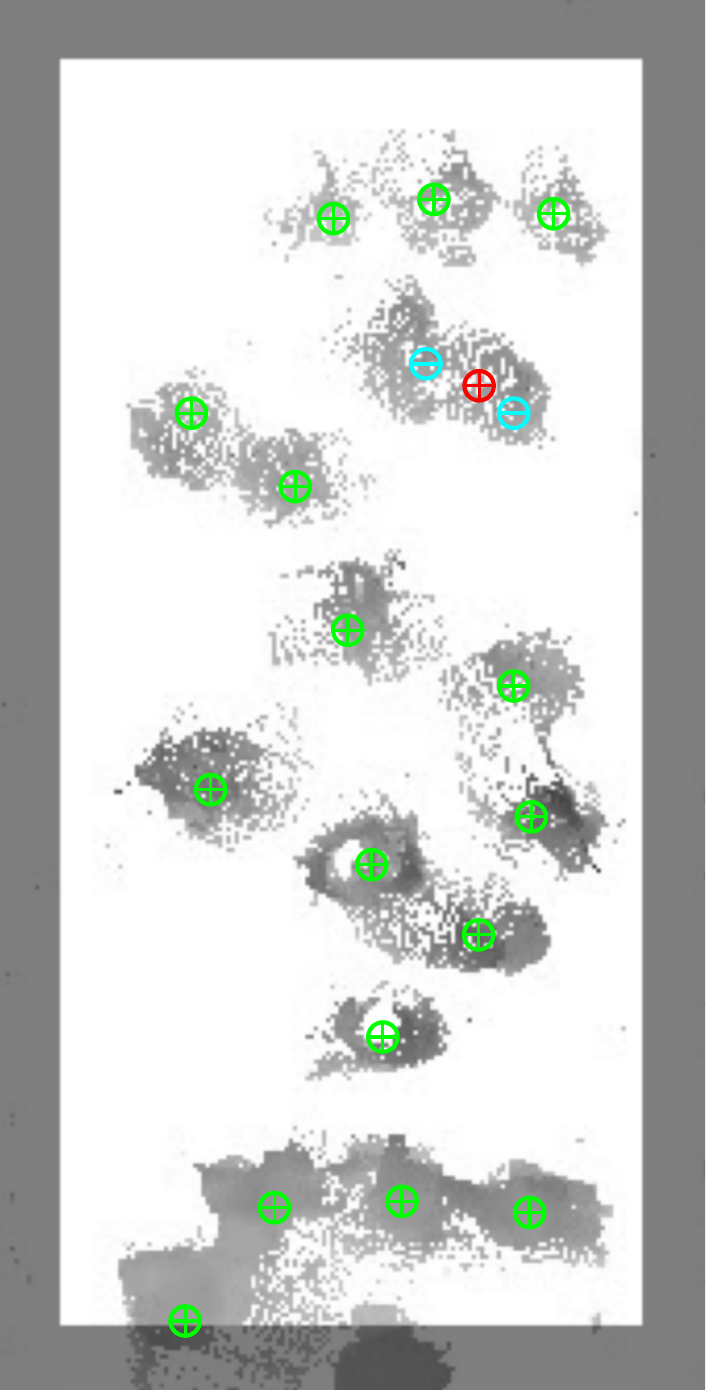} 
    \put(5, 92){(c)}
    \put(5, 97){\color{white}$0.72$ ped/m$^2$}
  \end{overpic}
  \begin{overpic}[width=.3\linewidth]{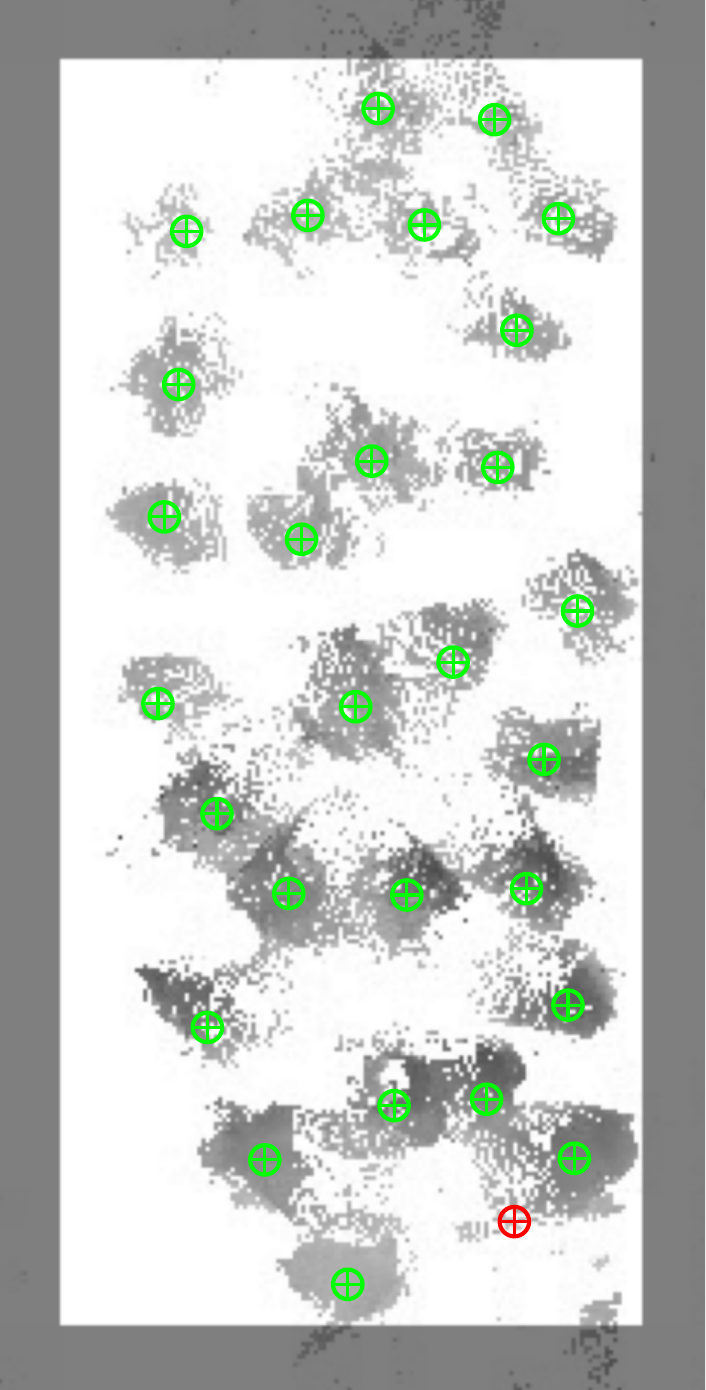} 
    \put(5, 92){(d)}
    \put(5, 97){\color{white}$1.12$ ped/m$^2$}
  \end{overpic}
  \begin{overpic}[width=.3\linewidth]{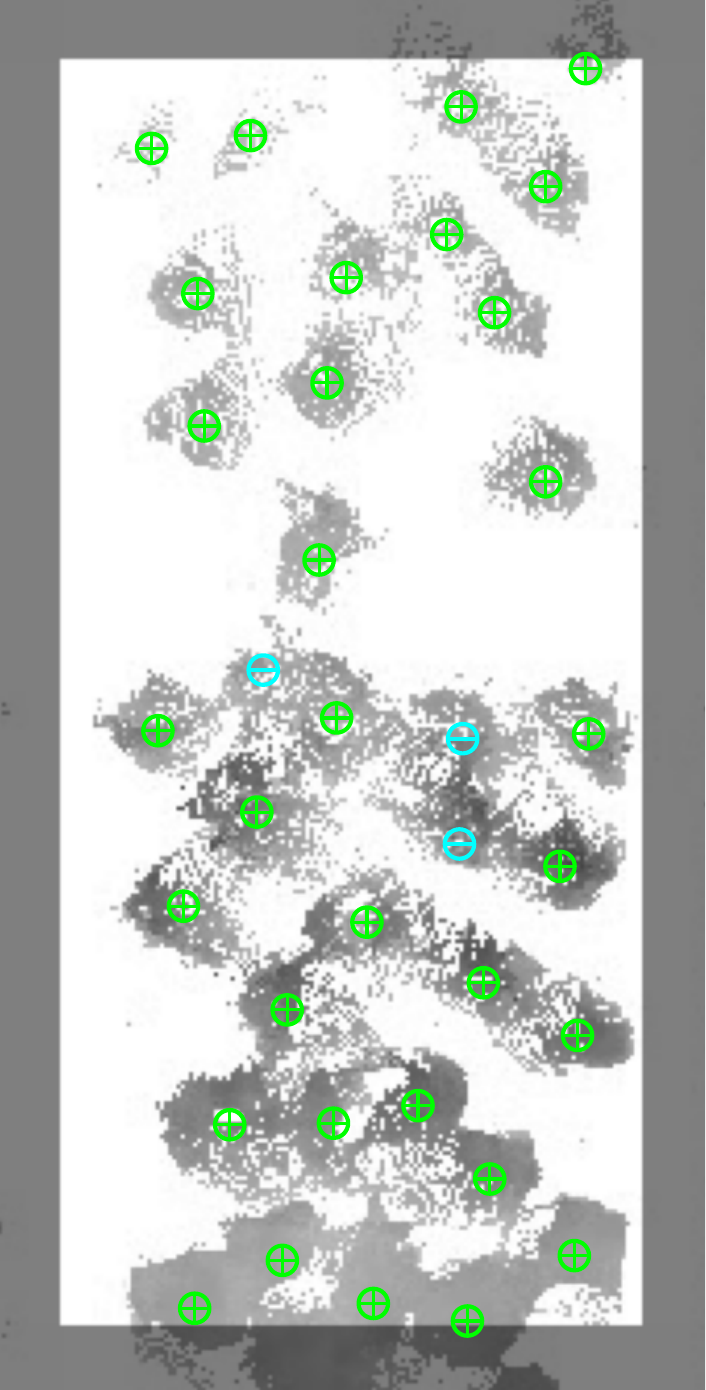} 
    \put(5, 92){(e)}
    \put(5, 97){\color{white}$1.40$ ped/m$^2$}
  \end{overpic}
  \begin{overpic}[width=.3\linewidth]{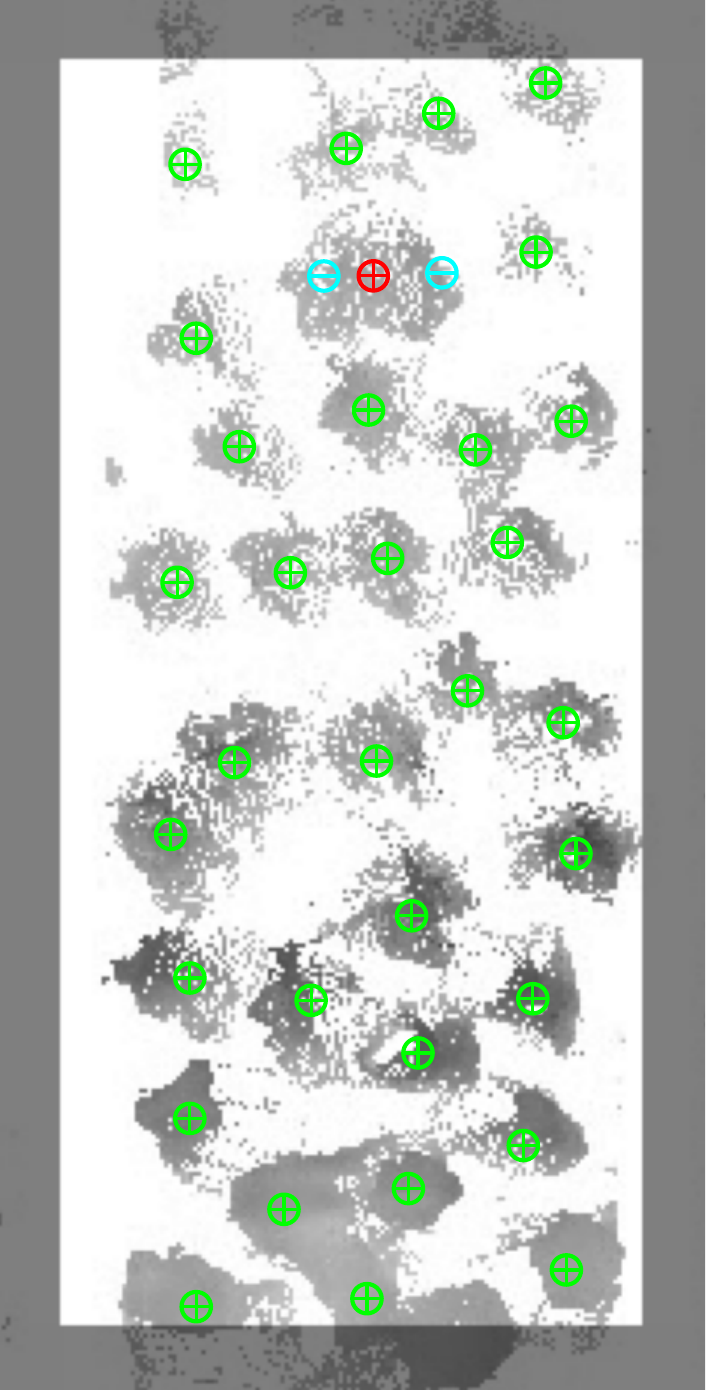} 
    \put(5, 92){(f)}
    \put(5, 97){\color{white}$1.36$ ped/m$^2$}
  \end{overpic}

  \begin{overpic}[width=0.5\linewidth]{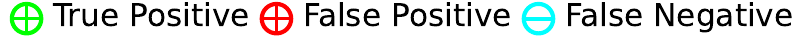} 
  \end{overpic}

  \caption{Typical corrections made during the model assisted annotations. Global density of the scenes are indicated at the top of each image. The bounding boxes are not displayed to improve readability. (a,d) Straightforward false positive. (b,e) Straightforward false negative. (c,f) Multiple pedestrians detected as a single pedestrian, resulting in a false positive and multiple false negatives.}\label{fig:dataset-corrections}
\end{figure}
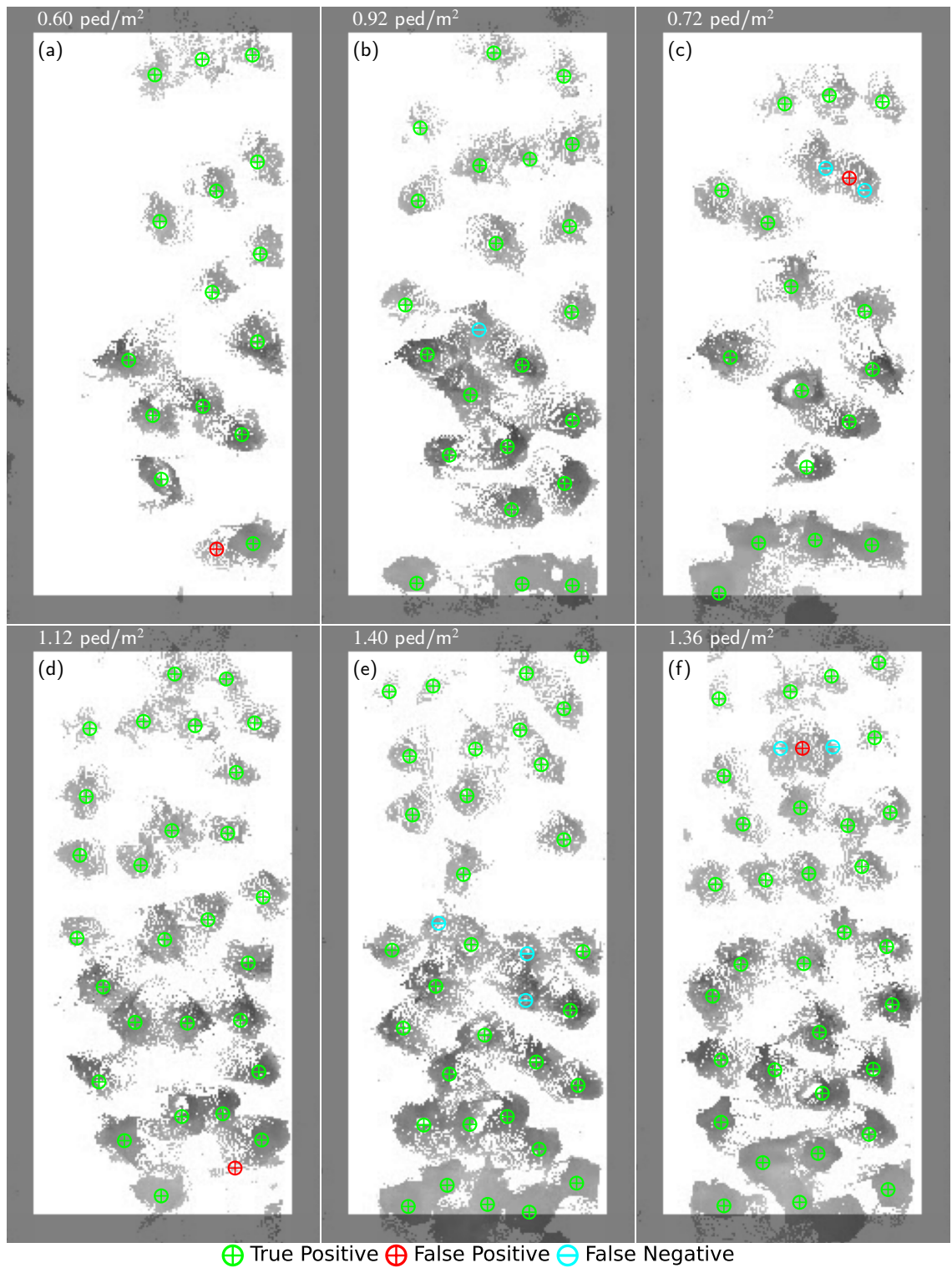

We utilize the resulting model to aid in the annotation of the next data set. The model generates initial annotations for new frames outside the existing dataset. We conduct a comprehensive review of these annotations, wherein we approve the correct annotations, add any missed annotations, and correct any erroneous annotations. The typical corrections that are made during the review process can be divided into three categories. The first category are straightforward false positives (Fig.~\ref{fig:dataset-corrections}a,d), in which the model incorrectly annotates a non-pedestrian as a pedestrian. The second category are straightforward false negatives (Fig.~\ref{fig:dataset-corrections}b,e), this involves instances where the model fails to detect pedestrians that are present in the frame. The final category represents a more intricate scenario where the model mistakenly identifies multiple pedestrians walking in proximity of each other as a single pedestrian (Fig.~\ref{fig:dataset-corrections}c,f), resulting in a false positive and multiple false negatives.

The iterative process of utilizing the previous model to assist in the creation of the next data set increases the efficiency at which we can create new annotations. We use the same CutMix-like approach to create the next dataset with the new annotations. A new model is trained on this dataset using transfer learning. We repeat this process until the model is sufficiently accurate. The final dataset consists of approximately $3700$ annotations distributed across $819$ images. Notably, this last dataset encompasses frames that capture a wider range of global density compared to the initial iteration of the data set.

\subsection{Pedestrian localization model validation}

During the validations process, we quantify the accuracy of the model considering $500$ frames outside the training dataset. To ensure a thorough evaluation, we select frames covering a diverse range of local densities. These frames are then divided into equally sized density bins, with each bin containing $100$ images. We conduct a thorough review of the model-generated annotations, following the same correction process as during the dataset generation. In this validation, any erroneous annotations are considered false positives, while additional annotations are considered false negatives. 

\begin{figure}[t]
  \centering
  \includegraphics[width=.6\linewidth]{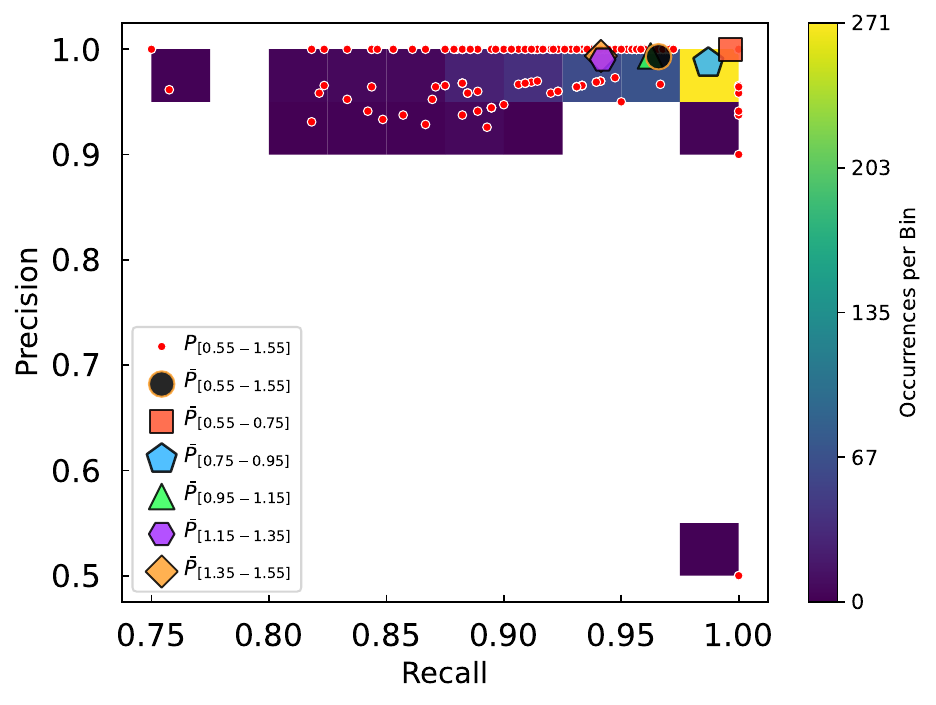} 
  \caption{Precision-recall curve ($P_{[0.55-1.55]}$) of our model, along with the mean precision-recall $\bar{P}_{[0.55-1.55]}$, and the number of occurrences per indicated precision-recall bin for the entire evaluated local density domain, ranging from $0.55$ ped/m$^2$ to $1.55$ ped/m$^2$ in $500$ frames. The plot also includes the means ($\bar{P}$) of the different evaluated density bins. Showcasing the exceptional accuracy of our model for densities below $0.95$ ped/m$^2$, while still performing commendably up to $1.55$ ped/m$^2$}\label{fig:yolo-accuracy}
\end{figure}

We assess the quality of the model by the precision ($P_e$), recall ($R_e$) and F1 score ($F_1$) of the model using the reviewed frames, defined as follows:
\begin{equation}\label{eq:fp-tp-f1}
  P_e = \frac{T_P}{T_P + F_P}, \quad R_e = \frac{T_P}{T_P + F_N}, \quad F_1 = 2 \frac{P_e R_e}{P_e + R_e},
\end{equation}
where $T_P$ is the number of true positives, $F_P$ is the number of false positives and $F_N$ is the number of false negatives. Precision indicates the fraction of correct detections with respect to the total number of detections, offering a measure of the quality of the detections. Recall denotes the fraction of correct detections with respect to the total number of expected detections, providing a measure of the quantity of relevant detections. The F1 score represents the harmonic mean of the precision and recall. In the ideal case of a perfect model all these parameters are equal to 1.
We visualize the quality of the model through a precision-recall curve in Fig.~\ref{fig:yolo-accuracy}. This curve, represented by ($P_{[0.55-1.55]}$), illustrates the relationship between recall and precision for the evaluated local density domain, which ranges from $0.55\;$ped/m$^2$ to $1.55\;$ped/m$^2$. Additionally, the mean precision-recall, $\bar{P}_{[0.55-1.55]}$, for this entire density domain is plotted, along with the means ($\bar{P}$) of the different density bins. Finally, the number of occurrences per indicated precision-recall bin is displayed in the plot, showcasing the significant number of frames with both good precision and good recall.

Fig.~\ref{fig:yolo-accuracy} and Table~\ref{tab:yolo} provide an overview of the performance of our model across various precision and recall levels. It demonstrates that our model exhibits exceptional accuracy for densities below $0.95\;$ped/m$^2$ and still performs extremely well for densities up to $1.55\;$ped/m$^2$, which is close to the maximum local density recorded. 

\section{Fundamental diagrams based on hydrodynamic density}\label{app:comparison}
\begin{figure}[t]
  \centering
  \begin{overpic}[width=.3\linewidth]{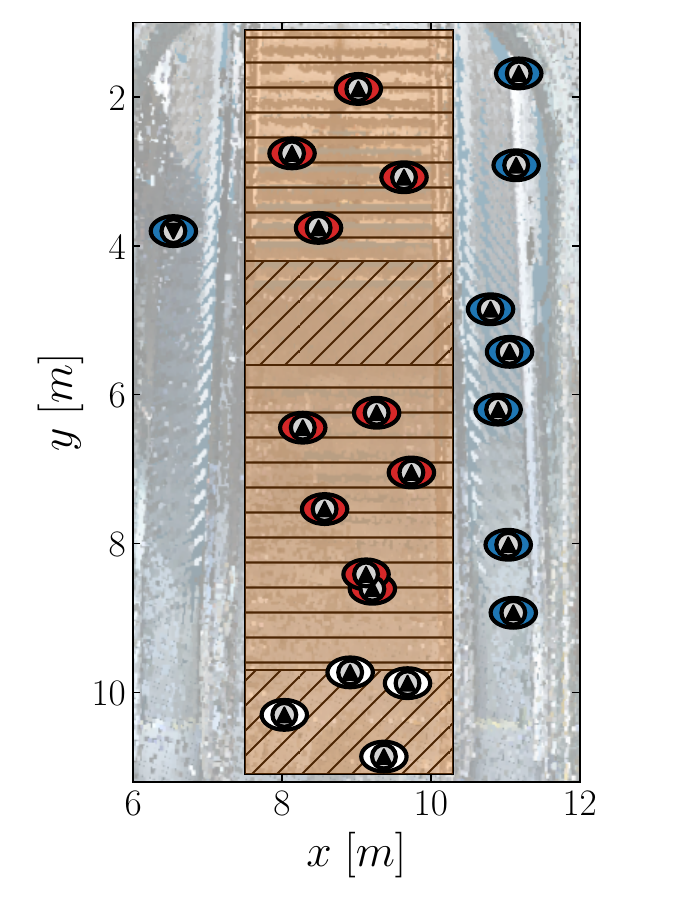} 
    \put(1, 100){(a)}
  \end{overpic}
  \begin{overpic}[width=.6\linewidth]{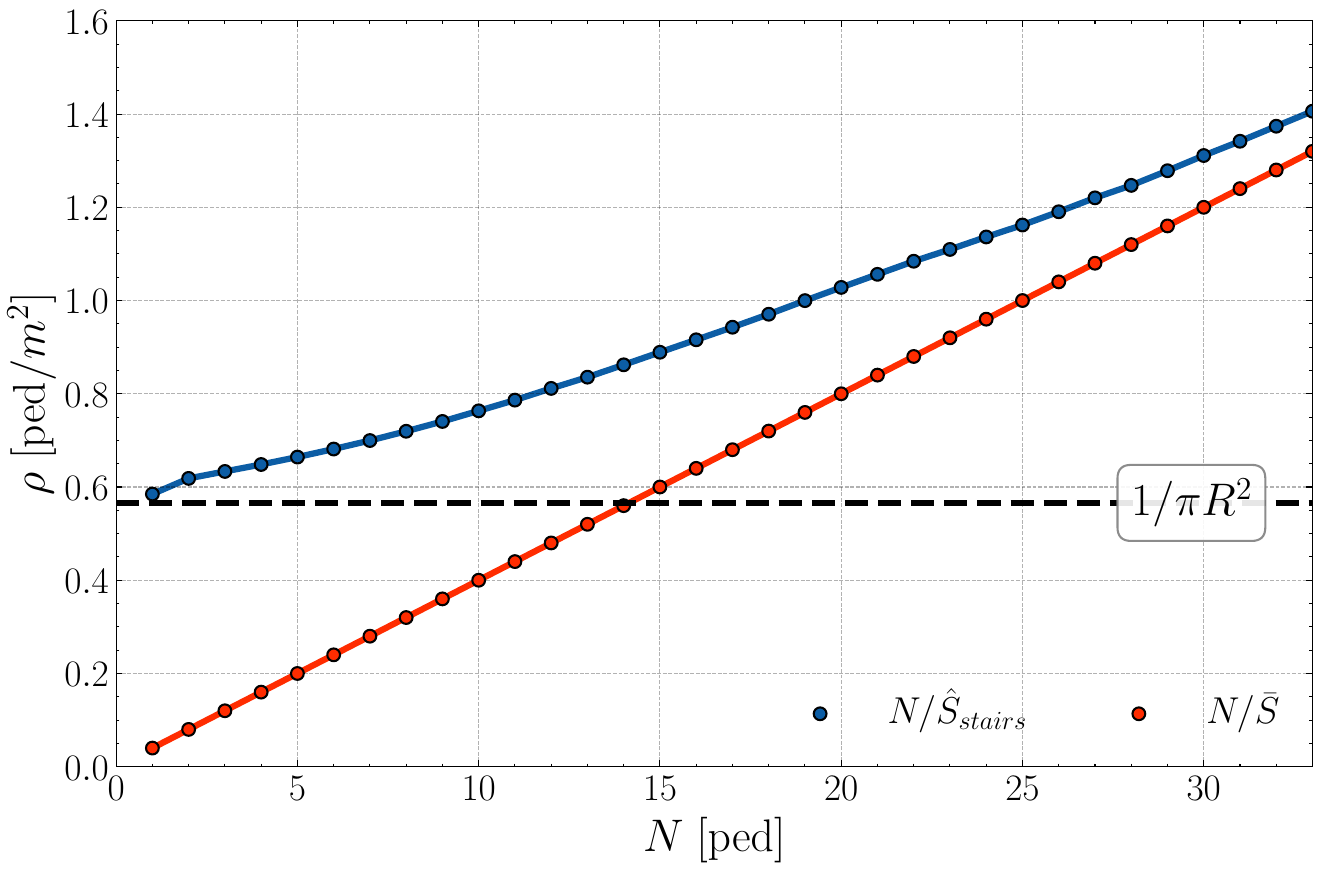} 
    \put(1, 67){(b)}
  \end{overpic}
   \begin{overpic}[width=\linewidth]{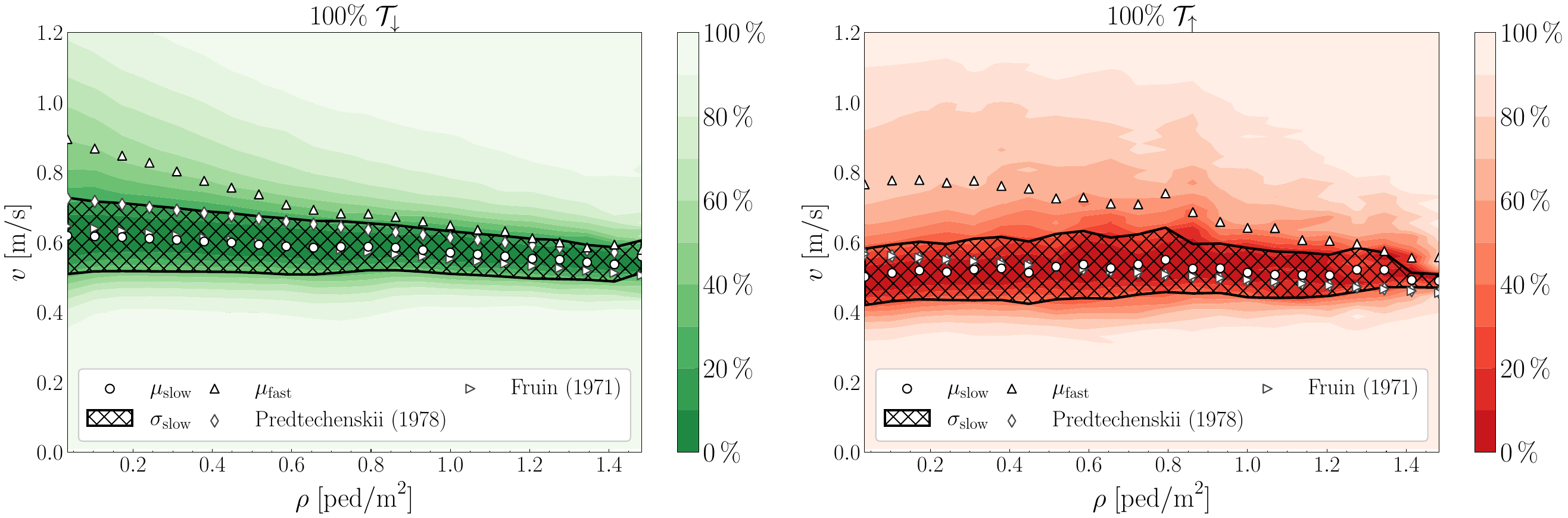} 
     \put(1, 33){(c)}
     \put(50, 33){(d)}
  \end{overpic}
  
  \caption{\label{fig:area-vs-density} Fundamental diagram using the
    classical method to compute the crowd density. (a) Illustration of
    a frame with, $N=14$ pedestrians. Global surface area used to
    calculate the density indicated with an orange color. (b)
    Comparison of the global density calculated using the classical
    method from Eq.~\eqref{eq:classical-density} (red dots) with
    $A=25\;$m$^2$, and the local density calculation method from
    Eq.~\eqref{eq:density-hat} (blue dots) (c) Classical fundamental
    diagram for unidirectional flow of descending pedestrians
    ,$100\% \;\mathcal{T}_{\downarrow}$, and (d) for ascending
    pedestrians, $100\% \;\mathcal{T}_{\uparrow}$.}
\end{figure}

Traditionally, the fundamental diagram is computed by employing the hydrodynamic definition from Eq.~\eqref{eq:classical-density} to calculate the pedestrian density. In this work however, we estimate the pedestrian density as the ratio between the number of pedestrians and the union of all the personal spaces as defined in Eq.~\ref{eq:density-hat} for reasons elaborated in Sec.~\ref{sec:density-vel}. This appendix provides a comparison between the classical method and the method employed in this work.
In Fig.~\ref{fig:area-vs-density}a we provide an example of a frame with $N=14$ pedestrians where the global area $\overline{S}=25\;$m$^2$ is indicated by the orange colored domain. This example yields a hydrodynamic density $\rho=14/25\approx 0.56\;$ped/m$^2$.
In Fig.~\ref{fig:area-vs-density}b we provide a comparison between the pedestrian density employed in this work $N/\hat{S}_{stairs}$ and the classical density $N/\overline{S}$ for increasing number of pedestrians $N_{stairs}$. We observe that our method has a lower bound determined by not intersecting personal spaces $1/(\pi R^2)=0.56\;$m$^2$. For increasing number of pedestrians we observe that our method converges to the classical approach.  
For comparison with the fundamental diagram from Fig.~\ref{fig:fd-ps} we report the classical fundamental diagram in
Fig.~\ref{fig:area-vs-density}c,d.

\printcredits
\newpage
\bibliographystyle{cas-model2-names}

\bibliography{biblio.bib}

\end{document}